\UseRawInputEncoding

\documentclass[10pt, prb, aps, twocolumn, showpacs, citeautoscript, floatfix, reprint, amsmath, amssymb, notitlepage,  superscriptaddress]{revtex4-1}

\usepackage{amssymb}
\usepackage{bm}
\usepackage{epsfig}
\usepackage{wasysym}
\usepackage{bbm}

\usepackage{color}
\newcommand{\cred}{\color{red}}
\newcommand{\cblue}{\color{blue}}

\usepackage{macros}
\renewcommand{\la}{$\leftarrow$}

\usepackage{braket}
\newcommand{\dhat}{{\vec {n}}}
\renewcommand{\k}{\vec k}
\newcommand{\trimk}{\vec{\overline k}}

\begin{document}

\title{Unification of  topological invariants in Dirac models}

\author{Gero von Gersdorff}
\affiliation{Department of Physics, PUC-Rio, 22451-900 Rio de Janeiro, Brazil}

\author{Shahram Panahiyan}
\affiliation{Helmholtz-Institut Jena, Fr\"{o}belstieg 3, D-07743 Jena, Germany}
\affiliation{Helmholtzzentrum f\"{u}r Schwerionenforschung, D-64291 Darmstadt, Germany}
\affiliation{Theoretisch-Physikalisches Institut, Friedrich-Schiller-University Jena, D-07743 Jena, Germany}

\author{Wei Chen}
\affiliation{Department of Physics, PUC-Rio, 22451-900 Rio de Janeiro, Brazil}

\date{\today}

\begin{abstract}

Topological phases of materials are characterized by topological invariants that are conventionally calculated by different means according to the dimension and symmetry class of the system. For topological materials described by Dirac models, we introduce a wrapping number as a unified approach to obtain the topological invariants in arbitrary dimensions and symmetry classes. Given a unit vector that parametrizes the momentum-dependence of the Dirac model, the wrapping number describes the degree of the map from the Brillouin zone torus to the sphere formed by the unit vector that we call Dirac sphere. This method is gauge-invariant and originates from the intrinsic features of the Dirac model, and moreover places all known topological invariants, such as Chern number, winding number, Pfaffian, etc, on equal footing.  

\end{abstract}

\maketitle

\section{Introduction}
\label{sec:intro}

The topological order in topological insulators (TIs) and superconductors (TSCs) is generally defined through the Bloch states according to the dimension and symmetry class of the system\cite{Schnyder08,Ryu10,Kitaev09,Chiu16}. Classification of these materials according to the time-reversal (TR), particle-hole (PH), and chiral symmetries renders a table of 10 symmetry classes in which the structure of the topological invariants repeats every 8 dimensions, often referred to as the periodic table of topological invariants. In a given symmetry class, the topological invariant (or charge) characterizes equivalence classes of different Hamiltonians that possess the same PH and TR symmetries.  Hamiltonians with the same charge are equivalent under certain continuous deformations \cite{Kitaev09} while those of unequal charges are not. 
A commonly employed quantity to calculate these invariants is the gauge connection 
\be
\mathcal A_\mu^{ab}(\vec k)\equiv\braket{\psi^a_-(\vec k)|\partial_\mu|\psi^b_-(\vec k)}
\ee
where the states run over all the filled-band  Bloch eigenstates, and $\vec k$ is the momentum within the first Brillouin Zone (BZ). A unitary ($\vec k$-dependent) change of basis  within the filled band eingenstates, generates a gauge transformation for the system.
The invariants for all known symmetry classes can then be formulated in terms of certain gauge-invariant functionals of $\mathcal A_\mu$ (usually integrals over the BZ), such as Chern-number, winding-number, Chern-Simons number, etc.\cite{Chiu16}  At times the functionals are defined only in certain classes of gauges governed by rather complicated constraints.
In general, the gauge dependence of the charges is not always manifest, and the additional constraints make some of them hard to compute in practice. A very useful overview of all the topological invariants in the gauge formalism can be found in Ref.~\onlinecite{Chiu16}.

The periodic table of  topological invariants has the following structure. Of the ten symmetry classes, eight of them are the "real" symmetry classes, which possess TR and/or PH symmetry, and the other two are the "complex" ones, which do not.
In particular, in each dimension $D$ there exists one 
nontrivial class among the eight real symmetry classes, which is different for each $0\leq D\leq 7$ and  repeats periodically thereafter. This so-called "primary series" is characterized by an integer topological invariant, which, according to whether $D$ is even or odd, can be given in terms of either the Chern or winding number.
Another similar series, the so-called "even series" is characterized by an even integer invariant.
The nontrivial complex classes have only periodicity two, they have also integer charges and we will refer to them as the complex series.
Starting from the primary series, one can obtain the remaining nontrivial real symmetry classes by considering the same Hamiltonians, but in one and two dimensions lower respectively.\cite{Qi08,Ryu10} These are  known as first and second descendants, and they are characterized by a binary invariant $\pm 1$. The binary charges can be computed in a more or less complicated construction that makes reference to the invariants of the parent (primary) class.

The goal of this paper is to provide a simple, practical, and unified characterization of the topological invariants in all nontrivial classes in any $D$ that is manifestly gauge invariant.
We will focus our attention on the Dirac models that describe most of the topological materials, and in the context of the K-theory approach \cite{Kitaev09} have been argued to provide a sufficiently general class of models.  Our formalism is based on the fact that Dirac Hamiltonians in $D$ dimensions can be considered as maps from the $D-$dimensional BZ torus $T^D$ to a $D-$dimensional sphere $S^D$ formed by the unit vector that parametrizes the Dirac Hamiltonian. For such maps, one can define the degree of the map which counts the number of times the BZ covers this sphere. We will call the degree of map the {\it wrapping number}. It can either be given as an integral or in terms of a rather simple algebraic formula, whose equivalence was shown by Kronecker over a century ago \cite{Kronecker1869}. 
We will show that in the primary and complex series, the topological charge is always given by the wrapping number (for any $D$, even or odd), in the even series it is given by twice the wrapping number, and in the case of the descendants it is given by $-1$ to the power of the wrapping number.
Moreover, using the algebraic formula, we will prove an extremely simple non-integral version of the binary charges for the case of Dirac Hamiltonians that also makes close connection to the known formulation of binary invariants in terms of Pfaffians.

\section{Unification of topological invariants \label{sec:quantum_metric_general}}

\subsection{Dirac Hamiltonians}
\label{sec:dirac}

We consider the TIs and TSCs classified according to the TR, PH and chiral symmetries, under which the first-quantized Hamiltonian $H(\vec k)$ transforms as
\begin{eqnarray}
&&TH^*({\bf k})T^{-1}=H(-{\bf k})\;,
\nonumber \\
&&CH^*({\bf k})C^{-1}=-H(-{\bf k})\;,
\nonumber \\
&&SH({\bf k})S^{-1}=-H({\bf k})\;.
\label{General_symmetries_TR_PH_CH}
\end{eqnarray}
Notice that in our convention $T$ and $C$ do not include complex conjugation (which is instead shown explicitly in Eq.~(\ref{General_symmetries_TR_PH_CH})). Presences of both $T$ and $C$ guarantees existence of $S$ via $S=TC$. 
This classification yields five topologically nontrivial classes in each dimension, as reviewed in section \ref{sec:intro}. We will focus on the Dirac Hamiltonians that realize TIs and TSCs according to their symmetry classes. To explicitly construct them, we start by defining the $SO(2n+1)$ Clifford algebra
\be
\{\Gamma_i,\Gamma_j\}=2\delta_{ij}\,,\qquad 1\leq i,j\leq  2n+1\,,
\ee
in which the $\Gamma_i$ are $N\times N$ dimensional, Hermitian matrices, where $N=2^n$. 
We will assume that our system is defined by a Bloch Hamiltonian that takes the form of a Dirac Hamiltonian
\be
H= \sum_{i=0}^Dr^i(\vec k)\Gamma_i,
\label{eq:dirac}
\ee
where $ \vec r$ is a vector and $k^\mu$ is the $D$ dimensional momentum, and $\Gamma_0$ depends on the symmetry class,  as detailed in App.~\ref{sec:details}.

For convenience we will work with the "spectrally flattened" Hamiltonian
\be
Q\equiv |\vec r|^{-1}H=\vec n\cdot\Gamma\,,\qquad 
\dhat \equiv \frac{\vec r}{|\vec r|}, 
\label{spectrally_flattened}
\ee
which has the same eigenstates as $H$ but its eigenvalues are $\pm 1$ instead of $\pm r$.
The  manifold described by the unit vector $\dhat$ is a $D$ dimensional sphere $S^D$ that we will refer to as the  {\it Dirac sphere} in what follows. Note that in the case of $N=2$, $D=2$, the Dirac sphere is naturally identified with the familiar Bloch sphere $\mathbb {CP}^1$.

\begin{figure}
\includegraphics[width=0.8\linewidth]{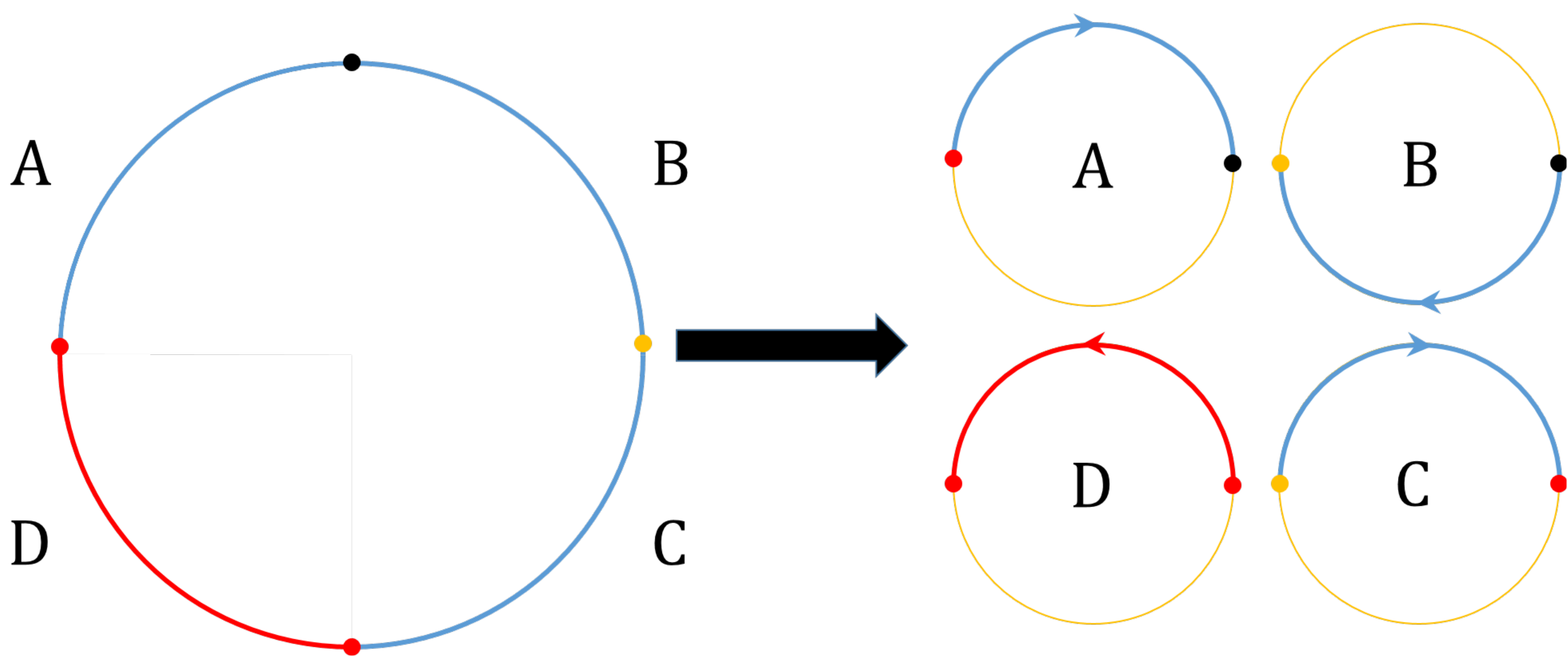}

\vspace{0.5cm}

\includegraphics[width=0.8\linewidth]{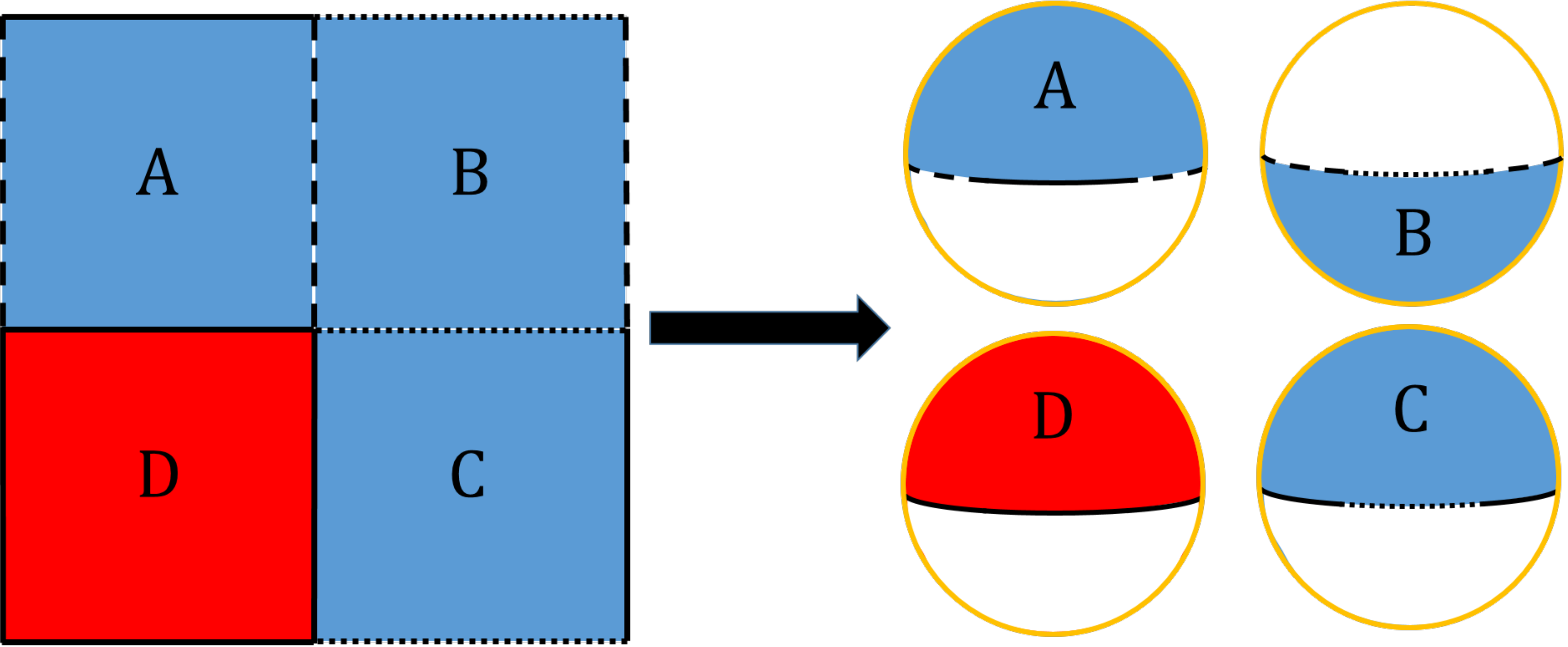}
\caption{Schematic examples of maps  from the BZ torus to the Dirac sphere, $\vec n(\vec k):T^D\to S^D$, with wrapping number $+1$, with $D=1$ ($D=2$) in the upper (lower) panel. In both cases we divide the domain $T^D$ into four regions (A,B,C,D) and separately show their images in $S^D$ on the right.
The blue color indicates points with positive Jacobian (orientation preserving) and the red one negative Jacobian (orientation reversing). As can be verified from the colors ($D=1$) and different dashed lines ($D=2$), the maps are indeed continuous.}
\label{fig:wrap2}
\end{figure}

\newcommand{\degnc}{{$\deg [\vec n]$}}
\newcommand{\degn}{{ $\deg [\vec n]$}}
\newcommand{\tdegn}{{ $2\deg [\vec n]$}}
\newcommand{\degt}{{ $(-1)^{\deg[\vec n]}$}}
\begin{table*}
{
\begin{center}
\begin{tabular}{c cccccccc}
			&$D=0$	&$D=1$	&$D=2$	&$D=3$	&$D=4$	&$D=5$	&$D=6$	&$D=7$\\
\hline
A			&\degnc	&		&\degnc	&		&\degnc	&		&\degnc	&\\
AIII		&		&\degnc	&		&\degnc	&		&\degnc	&		&\degnc		\\		
\hline
AI			&\degn	&		&		&		&\tdegn	&		&\degt	&\degt\\
BDI			&\degt	&\degn	&		&		&		&\tdegn	&		&\degt\\
D 			&\degt	&\degt	&\degn	&		&		&		&\tdegn\\
DIII		&		&\degt	&\degt	&\degn	&		&		&		&\tdegn\\
AII			&\tdegn	&		&\degt	&\degt	&\degn	&		&		\\
CII			&		&\tdegn	&		&\degt	&\degt	&\degn\\
C 			&		&		&\tdegn	&		&\degt	&\degt	&\degn\\
CI 			&		&		&		&\tdegn	&		&\degt	&\degt	&\degn
\end{tabular}
\end{center}
}
\caption{  Unification of
topological invariants in the periodic table of topological materials. Given a Dirac Hamiltonian $H(\vec k)=\sum_{i=0}^D r_i(\vec k) \Gamma_i$ that describes a topologically nontrivial system in certain symmetry class and dimension, $\deg [\vec n]$ denotes the 
degree of the map $\vec n(\vec k): T^D\to S^D$  from the BZ torus to the Dirac sphere given by the $\vec n=\vec r/|\vec r|$ vector. This degree of the map is referred to as the wrapping number, defined explicitly in Eq.~(\ref{eq:deg}). The table continues periodically for $D>7$. 
}
\label{tab:invariants}
\end{table*}

\subsection{The wrapping number}
\label{sec:wrapping}

We will now define the central quantity of our paper  that in differential geometry is usually called the degree of the map $\dhat(\vec k)$
\bea
\deg [\dhat]&\equiv& \frac{1}{V_D}\int_{BZ} \frac{1}{D!}\eps_{i_0\cdots i_D}n^{i_0}dn^{i_1}\wedge\dots \wedge dn^{i_D},\nn\\
&=&\frac{1}{V_D}\int  \eps_{i_0\cdots i_D}n^{i_0}\partial_1 n^{i_1}\dots  \partial_D n^{i_D}\,d^{D}k\,,\nn\\
&=&\frac{1}{V_D}\int  \eps_{i_0\cdots i_D}\frac{1}{|\vec r|^{D+1}}r^{i_0}\partial_1 r^{i_1}\dots  \partial_D r^{i_D}\,d^{D}k\,,\;\;\;\;
\label{eq:deg}
\eea
where $V_D$ is the volume of the $D$ dimensional unit sphere,
\be
V_D=\frac{2\pi^{\frac{D+1}{2}}}{\Gamma(\frac{D+1}{2})}.
\label{eq:SDsphere}
\ee
The quantity $\deg[\vec n]$  is always an integer and counts how many times the domain manifold $T^D$ wraps around the image manifold $S^D$ under the map $\dhat(k)$. We will refer to $\deg[\vec n]$ as the wrapping number. 
Eq.~(\ref{eq:deg}) is a special case of a more general definition of the degree of a map between two arbitrary manifolds of the same dimension, which we review in App.~\ref{sec:deg}.
As we will see, in any dimension, all known topological invariants such as Chern numbers, winding numbers, Chern-Simons-invariants, Pfaffians, etc.~can be expressed in terms of $\deg[\vec n]$. In Fig.~\ref{fig:wrap2}, a 1D and a 2D  schematic example of wrapping number $\deg[\vec n]=+1$ are given.

The degree of the map can be expressed in an alternative way which turns out to be extremely useful for the practical computation of topological invariants. 
The method goes as follows: pick any point on the sphere $\vec n_0$, and find all points $\vec k_i$ in the domain that map to this point, $\dhat(\vec k_i)=\vec n_0$. If the point is chosen such that the "Jacobians" at points $\vec k_i$
\be
J_{\vec n}(\vec k_i)\equiv \left.\eps_{i_0\cdots i_D}n^{i_0}\frac{\partial n^{i_1}}{\partial k^1}\cdots \frac{\partial n^{i_D}}{\partial k^D}\right|_{\vec k_i},
\ee
are all nonvanishing, then the set of these points is discrete and finite, and 
\be
\deg[\dhat]=\sum_{\substack{\vec k\ {\rm with}\\ \vec n(\vec k)=\vec n_0}} \operatorname{sign}J_{\vec n}(\vec k),
\label{eq:discrete}
\ee 
which is manifestly an integer. The equivalence of Eq.~(\ref{eq:deg}) and Eq.~(\ref{eq:discrete}) was proven in a slightly different form by Kronecker\cite{Kronecker1869} and later generalized by Brouwer\cite{Brouwer1912}. 
Finally, since $J_{\vec r}=|\vec r|^{D+1}J_{\vec n}$ according to Eq.~(\ref{eq:deg}), one can also compute $\operatorname {sign} J_{\vec n}=
\operatorname {sign} J_{\vec r}$ from the cyclic derivative of the ${\bf r}$ vector, which is often much simpler.


In subsections \ref{sec:winding} to \ref{sec:des}, we will show that the nontrivial Altland-Zirnbauer (AZ) classes\cite{Altland97} represented by Dirac Hamiltonains can be classified according to the wrapping number, as  summarized in Table \ref{tab:invariants}. It should also be emphasized that we do not set out to rederive the periodic table of topological materials from scratch, but rather construct our formalism based the existing knowledge, in particular Ref.~\onlinecite{Ryu10} and \onlinecite{Chiu16}, and only focus on those classes and dimensions that are topologically nontrivial. We furthermore stress a very important property of the wrapping number,  namely it is manifestly gauge-invariant, where the gauge means the $U(N/2)$ freedom of choosing a basis for the filled-band eigenstates. 
 In fact, the wrapping number can be viewed as an intrinsic property of the Hamiltonian.

We also remark that the wrapping number is in accordance with a unified description of topological phase transitions (TPTs) based on the assumption that the topological invariants are generally momentum space integration of a certain curvature function, whose critical behavior gives rises to the statistical aspects of critical exponents, scaling laws, correlation functions\cite{Chen17,Chen19_AMS_review,Molignini18_Floquet_Majorana,
Molignini20_multicritical,Rufo19,Abdulla20,Kumar20}, universality classes\cite{Chen19}, fidelity susceptibilty\cite{Panahiyan20_fidelity_susceptibility,Molignini21_Kitaev_cross_dim}, and renormalization group approach\cite{Chen16,Chen16_2}. Within this wrapping number formalism, the curvature function is simply the cyclic derivative of the ${\bf n}$-vector in Eq.~(\ref{eq:deg}). Precisely how these statistical aspects manifest within the context of the wrapping number is a fundamental issue that should be addressed elsewhere.

\begin{figure}
\includegraphics[width=0.9\linewidth]{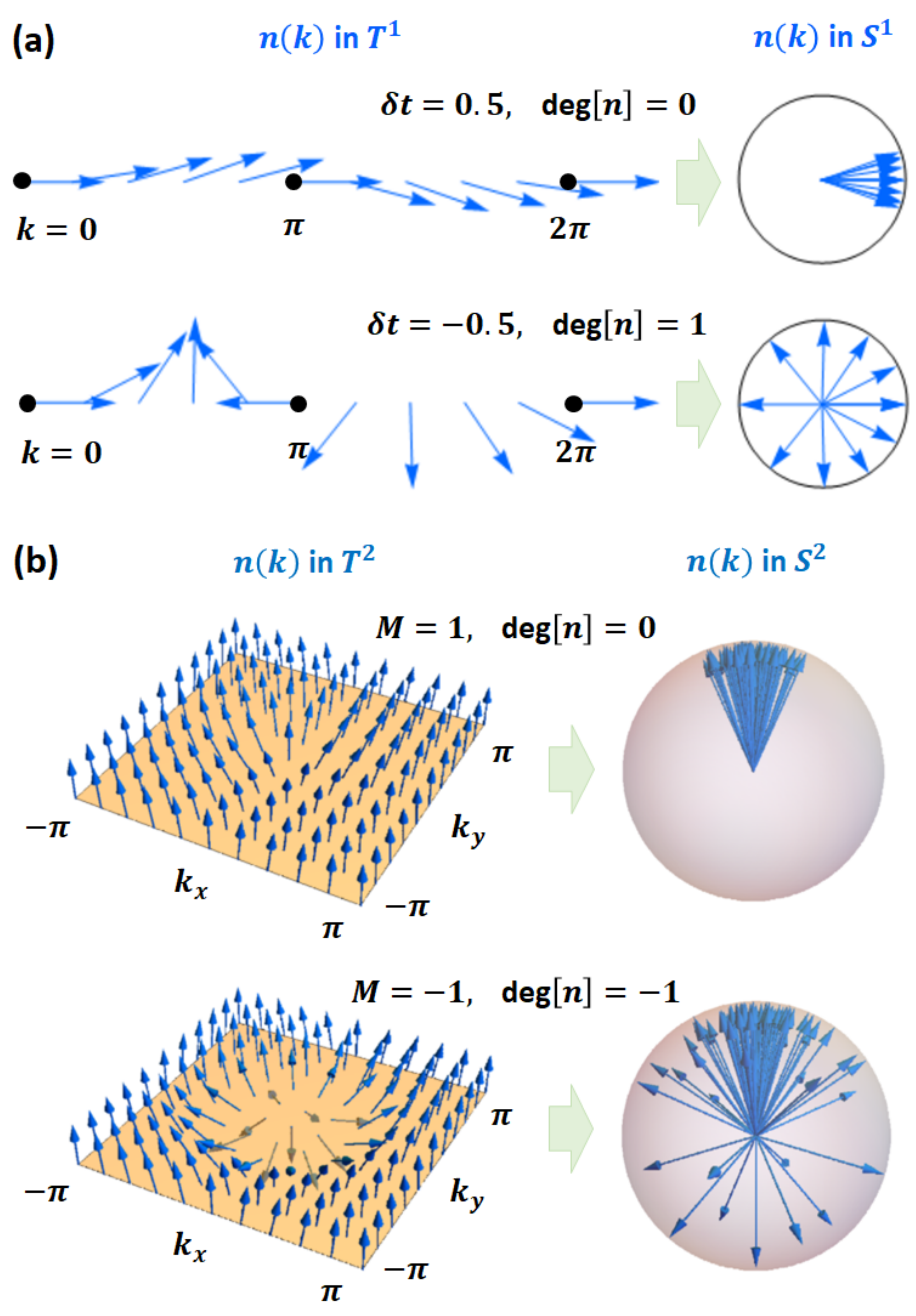}
\caption{(a) The ${\bf n}({\bf k})$ profile of the 1D SSH model in the BZ torus $T^{1}$, and in the Dirac sphere $S^{1}$ formed by shrinking the base of all the ${\bf n}({\bf k})$ vectors to the same point, in the topologically trivial $\delta t=0.5$ and nontrivial $\delta t=-0.5$ phases. (b) The ${\bf n}({\bf k})$ profile of the Chern insulator in the BZ torus $T^{2}$ and the Dirac sphere $S^{2}$, in the topologically trivial $M=1$ and nontrivial $M=-1$ phases. }
\label{fig:degn_2D_Chern}
\end{figure}

We proceed to use the prototype Su-Schrieffer-Heeger (SSH) model\cite{Su79} as a concrete 1D example to demonstrate the validity of Eqs.~(\ref{eq:deg}) and (\ref{eq:discrete}). The model is described by the lattice Hamiltonian\begin{eqnarray}
H&=&\sum_{i}\left[(t+\delta t)c_{Ai}^{\dag}c_{Bi}+(t-\delta t)c_{Ai+1}^{\dag}c_{Bi}+h.c\right],
\label{SSH_Hamiltonian}
\end{eqnarray}
where the different hoppings $t+\delta t$ and $t-\delta t$ on the even and odd bonds naturally distinguish the two sublattices $A$ and $B$. 
The $2\times 2$ Hamiltonian in momentum space takes the Dirac form $H(k)=r^{0}\sigma_{x}+r^{1}\sigma_{y}$, where $\sigma_{x}$ and $\sigma_{y}$ are Pauli matrices, with
\begin{eqnarray}
r^{0}=(t+\delta t)+(t-\delta t)\cos k,\;\;\;r^{1}=(t-\delta t)\sin k.\;\;\;
\end{eqnarray}
The profiles of the ${\bf n}(k)={\bf r}(k)/|{\bf r}(k)|$ vector in the BZ torus $T^{1}$ and the Dirac sphere $S^{1}$ are shown in Fig.~\ref{fig:degn_2D_Chern} (a). Choosing $\vec n_0=(1,0)$, for $\delta t/t>0$ the source points that map onto $\vec n_0$ are $k=0$ and $\pi$, whose $\operatorname{sign}J_{\vec n}(k)$ are $+$ and $-$, respectively, yielding $\deg[\dhat]=0$ according to Eq.~(\ref{eq:discrete}). For $\delta t/t<0$, there is only one source point $k=0$ with $\operatorname{sign}J_{\vec n}(k)=+$, yielding $\deg[\dhat]=1$. These results are consistent with the well-known fact that $\delta t/t<0$ is topologically nontrivial and $\delta t/t>0$ is trivial in this model. For any other choices of $\vec n_0$, one arrives at the same conclusion.

To demonstrate the same procedure in 2D, we consider the lattice model of Chern insulator described by the Hamiltonian $H({\bf k})=r^{0}\sigma_{z}+r^{1}\sigma_{x}+r^{2}\sigma_{y}$ with\cite{Bernevig13} 
\begin{eqnarray}
&&r^{0}=M+4-2\cos k_{x}-2\cos k_{y},
\nonumber \\
&&r^{1}=\sin k_{x},\;\;\;r^{2}=\sin k_{y},
\end{eqnarray}
whose ${\bf n}({\bf k})$ profile in the BZ torus $T^{2}$ and the Dirac sphere $S^{2}$ are shown in Fig.~\ref{fig:degn_2D_Chern} (b). Choosing ${\bf n}_{0}=(1,0,0)$, for $M>0$ the source points that have ${\bf n}_{0}$ are ${\bf k}=(0,0)$, $(0,\pi)$, $(\pi,0)$, and $(\pi,\pi)$, whose $\operatorname{sign}J_{\vec n}({\bf k})$ are $+$, $-$, $-$, $+$, rendering $\deg[\dhat]=0$. For $-2<M<0$, the source points are ${\bf k}=(0,\pi)$, $(\pi,0)$ and $(\pi,\pi)$, whose $\operatorname{sign}J_{\vec n}({\bf k})$ are $-$, $-$, $+$, corresponding to $\deg[\dhat]=-1$. As one can easily verify that $\deg[\dhat]$ is independent from the choice of ${\bf n}_{0}$, and correctly captures the well-known topological invariant in this model.

\subsection{Winding number in terms of wrapping number}
\label{sec:winding}

In odd dimensions, the systems 
in the primary, complex and even series have chiral symmetry, and
the topological invariant can be defined by the so-called winding number \cite{Chiu16}
\be
\nu_D = \frac{
(\frac{D-1}{2})!}{2D!} \left(\frac{i}{2\pi}\right)^\frac{D+1}{2}\int \tr S(QdQ)^D,
\ee
where $S$ is the generator of chiral symmetry (taken to be Hermitian) that satisfies $SH=-HS$, $Q$ is the spectrally flattened Hamiltonian $Q=\dhat\cdot \Gamma$  in Eq.~(\ref{spectrally_flattened}), and the power is to be understood as wedge products such that we obtain a $D$-form to integrate over.\footnote{In a basis where $ S=\Gamma_{2n+1}$ the Hamiltonian is off block diagonal and the winding number can be written in terms of the off-diagonal block matrices of dimension $N/2$.}
Using the definition of the volume of the $D$-dimensional sphere Eq.~(\ref{eq:SDsphere}) as well as $(QdQ)^2=-(dQ)^2$, this can be equivalently rewritten as
 \be
\nu_D  
=-\frac{1}{V_DD!} \left(-\frac{i}{2}\right)^\frac{D+1}{2}\int \tr [ SQ(dQ)^D].
\ee
It is not hard to see that this is, in fact, proportional to the wrapping number in the cases of the primary, even, and complex series:
\be
\nu_D =  2^{-\frac{D+1}{2}}\tr \left[{\mathbbm 1}\right] \, \deg[\dhat].
\ee
The proof of this formula uses the fact that 
\bea
&&\tr\left[SQ(dQ)^D\right]\nn\\
&=&\tr\left[S\Gamma_{i_0}\cdots\Gamma_{i_D}\right]n^{i_0}dn^{i_1}\wedge\cdots\wedge dn^{i_D}\nn\\
&=&\tr\left[S\Gamma_{0}\cdots\Gamma_{D}\right]\epsilon_{i_0\dots i_D}  n^{i_0}dn^{i_1}\wedge\cdots\wedge dn^{i_D}\nn\\
&=&-i^\frac{D+1}{2}\tr[{\mathbbm 1}]\epsilon_{i_0\dots i_D}  n^{i_0}dn^{i_1}\wedge\cdots\wedge dn^{i_D},
\eea
where in the last line we used Eq.~(\ref{eq:pseudoodd}).
Using the value $n=\frac{D+1}{2}$ for the primary and complex series one finds exactly
\be
\nu_D= \deg[\dhat].
\ee
Notable examples for this case include 1D class BDI and AIII, in which $\nu_{D}$ describes the quantized charge polarization of Wannier states\cite{KingSmith93,Resta94}, the 3D winding numbers for class AIII\cite{Schnyder08}, and the 3D class DIII realized in TR-invariant TSCs\cite{Fu10}. On the other hand, the value $n=\frac{D+3}{2}$ for the even series yields
\be
\nu_D 
=2\deg[\dhat],
\ee
and hence the winding number is twice the wrapping number.  This situation corresponds to 1D class CII that is relevant to several theoretical models\cite{Zhao14,Malard20_multicritical}, as well as 3D class CI.



\subsection{Chern number in terms of wrapping number}
\label{sec:chern}

In even dimensions, the systems in the  primary, complex and even series lack chiral symmetry, and the winding number is not defined. Instead, the topological invariant for these systems can be defined by the Chern number\cite{Chiu16}
\be
{\rm Ch}_\frac{D}{2}
=\frac{1}{V_DD!}\left(-\frac{i}{2}\right)^\frac{D}{2}\int \tr Q(dQ)^D.
\ee
Similar to the winding number, it is straightforward to express this as the wrapping number.
The calculation is very similar to the winding number. One has:
\bea
&&\tr\left[Q(dQ)^D\right]\nn\\
&=&\tr\left[\Gamma_{i_0}\cdots\Gamma_{i_D}\right]n^{i_0}dn^{i_1}\wedge\cdots\wedge dn^{i_D}\nn\\
&=&\tr\left[\Gamma_{0}\cdots\Gamma_{D}\right]\epsilon_{i_0\dots i_D}  n^{i_0}dn^{i_1}\wedge\cdots\wedge dn^{i_D}\nn\\
&=&i^\frac{D}{2}\tr[{\mathbbm 1}]\epsilon_{i_0\dots i_D}  n^{i_0}dn^{i_1}\wedge\cdots\wedge dn^{i_D},
\eea
where Eq.~(\ref{eq:pseudoeven}) was used.
The result, again valid for the primary, complex and even series, reads
\be
{\rm Ch}_\frac{D}{2}=2^{-\frac{D}{2}}\tr[{\mathbbm 1}]\deg[\vec n],
\ee

In the primary and complex series, one has $n=\frac{D}{2}$ and hence
\be
{\rm Ch}_\frac{D}{2}=\deg[\vec n].
\ee
This situation includes 2D class A that is relevant to a variety of important physical phenomena such as quantum Hall effect (QHE)\cite{Thouless82,Niu85}, quantum anomalous Hall effect (QAHE)\cite{Haldane88,Liu16}, Berry phase in TR-breaking systems\cite{Xiao05,Xiao10}, and the orbital magnetization\cite{Thonhauser05,Ceresoli06,Shi07,Souza08}. Besides, this situation also covers the 2D class D realized by chiral $p$-wave superconductors\cite{Read00}. For the even series, one has $n=\frac{D+2}{2}$ and hence
\be
{\rm Ch}_\frac{D}{2}=2 \deg[\vec n]\,,
\ee
and the Chern number is twice the wrapping number,  which is realized in 2D class C.


 We now comment on the peculiar case of class C. Notice that here we have implemented this class (which belongs to the even series) with $n=2$.
However, in this class in 2D one can in fact also choose $n=1$, with the following assignments
\be
\Gamma_{1,2}=\sigma_{1,2},\qquad \Gamma_0=\sigma_3,\qquad C=\sigma_2\,.
\ee 
The difference to the $n=2$ Dirac model is that all the parities are now positive, i.e.~$\vec r(-\vec k)=\vec r(\vec k)$.
For this model one has that ${\rm Ch}_{1}=\deg[\vec n]$ which naively can be even or odd. 
However, we can apply our expression Eq.~(\ref{eq:discrete})  to show that the wrapping number itself must be even in this case: For any point $\vec k$ with $\dhat (\vec k)=\dhat_0$, we have that also $\dhat (-\vec k)=\dhat_0$, and (as a consequence of the evenness of $D$) the Jacobians at these points are also equal. The only points where this doubling does not occur are the  high symmetry points ($\trimk =-\trimk$ modulo the dual lattice). However, it is also easy to see that at these points the Jacobian vanishes, and thus we must choose $\dhat_0\neq \vec n(\trimk)$ in order for Eq.~(\ref{eq:discrete}) to be well defined in this particular example. As just argued, at such $\dhat_0$ the source points come in pairs with equal Jacobians, and hence the wrapping number is even.

\subsection{Descendants}

\label{sec:des}

 Starting from each of the symmetry classes in the primary series, one can define so-called first and second descendants in the same symmetry class which live in one and two dimensions lower \cite{Qi08,Ryu10}, and whose invariants are binary, that is, an element of $\mathbb Z_2$ instead of $\mathbb Z$. 
For the remainder of this section, we denote  the dimension of the parent by $D$, and consequently the first descendant lives in $D-1$  and the second in $D-2$ dimensions.

As before, the spectrally flattened Hamiltonians are characterized by a map 
$\vec n(\k)$. Since the dimensionality of momentum space is reduced, the maps are of the type 
\bea
\vec n(\vec k)&:& T^{D-1}\to S^D\,,\qquad 1^{\rm st}\ \rm descendant
\label{eq:first}\\
\vec n(\vec k)&:& T^{D-2}\to S^D\,,\qquad 2^{\rm nd}\ \rm descendant
\label{eq:second}
\eea
The reality constraints coming from $T$ and $C$ transformations imply that the parities of the parent class satisfy
\be
\vec n(\vec k)=P\vec n(-\vec k)\,,\qquad 
P\equiv \diag(+1,-1,\dots,-1),
\label{eq:paritiesdes}
\ee
(see App.~\ref{sec:details} for details).
In practice, in Dirac models, the image of $\vec n(\vec k)$ lies in the lower dimensional sphere $S^{D-1}$ ($S^{D-2}$), that is, one (two) of the components of $\vec n(\vec k)$  are identically zero.~\footnote{However, the $C$ and $T$ symmetries do not enforce this  and the generic Dirac Hamiltonian of the given symmetry class is defined by Eq.~(\ref{eq:first}) and Eq.~(\ref{eq:second}) together with Eq.~(\ref{eq:paritiesdes}).} Then the maps $\vec n(\vec k)$ are of the type
$ T^{D-1}\to S^{D-1}$ ($T^{D-2}\to S^{D-2}$) for first (second) descendants.
In this case, the degree of the map is well defined and we will show that the binary invariants can be expressed in terms of them simply as
\be
P_{(1,2)}[\vec n]=(-1)^{\deg [\vec n]}.
\label{eq:des12}
\ee
We will also show (see Eq.~(\ref{eq:P1}) and Eq.~(\ref{eq:second})) that by means of Eq.~(\ref{eq:discrete}) one can derive an explicit and very simple formula for these invariants,
\be
P_{(1,2)}[\vec n]=\prod_\trimk n^0(\trimk),
\label{eq:P12}
\ee
 where $\trimk$ are the high symmetry points (HSPs) in momentum space that satisfies $\trimk=-\trimk$ up to a reciprocal lattice vector. Notice that because of parities, $n^i(\trimk)=0$ for $i\neq0$, and hence $n^0(\trimk)=\pm 1$. The construction of these invariants in our context follows closely the analysis of Refs.~\onlinecite{Qi08,Ryu10}, and we will show it in details in the following two subsections.

\subsection{First descendants}

\label{sec:firstd}

The spectrally flattened Hamiltonians are characterized by a map  given in Eq.~(\ref{eq:first}) with the constraint Eq.~(\ref{eq:paritiesdes}).
Let us consider two such Hamiltonians in a given symmetry class, parametrized by $\vec n_1(\vec k)$ and $\vec n_2(\vec k)$ and  define a continuous path $\dhat_{1\to 2}(\vec k,t)$ that connect these two,
\be
\dhat_{1\to 2}(\vec k,0)=\dhat_1(\vec k),\qquad 
\dhat_{1\to 2}(\vec k,\pi)=\dhat_2(\vec k). 
\ee 
We would like to interprete this path as a Hamiltonian in the parent class, in other words, we would like to view $t$ as an additional component of momentum. 
To comply with the relation $\dhat_{1\to 2}(-\k,-t)=P\dhat_{1\to 2}(\k,t)$ we need to extend the path to negative $t$ by 
\be
 \dhat_{1\to 2}(\k, t)=P \dhat_{1\to 2}(-\vec k,-t), \qquad {\rm for \ }-\pi<t\leq 0\,.
\ee
In this way, the newly defined $\dhat_{1\to 2}$ is a member of the parent class (the same symmetry class in $D$ dimensions).\footnote{It is maybe worth pointing out that for fixed times $t\neq 0,\pi$ the system does not belong to the $D-1$ dimensional symmetry class, so the interpolation is not to be regarded as a homotopy within the space of $D-1$ dimensional Hamiltonians of that class.}
Since the newly defined $\dhat_{1\to 2}$ is now a map form $T^{D}\to S^{D}$, one can compute the wrapping number of the parent class,
$
\deg[\dhat_{1\to 2}]
$
which is an integer. 
In principle, this integer can depend on the chosen path, however as we will show below, by choosing a different path it can only change by an even integer. Let us then define
\be
\deg_2[\dhat_{1\to 2}]\equiv \deg[\dhat_{1\to 2}]\mod 2.
\label{eq:ansatzZ2}
\ee
We thus declare two Hamiltonians topologically equal if $\deg_2[\dhat_{1\to 2}]=0$ and topologically different if $\deg_2[\dhat_{1\to 2}]=1$. This is the strategy employed in Refs.~\onlinecite{Qi08,Ryu10}, 
schematically, 
\be
\mathbb Z_2{\rm \ invariant}={\rm deg}[{\bf n}]({\rm interpolation})\mod 2
\ee
where ${\rm deg}[{\bf n}]$ here stands for Chern number\cite{Qi08} or winding number\cite{Ryu10} that is proved to be equal to the degree of the map in Secs.~\ref{sec:winding} and \ref{sec:chern}, leading us to consider directly Eq.~(\ref{eq:ansatzZ2}).


It remains to be shown that $\deg_2[\dhat_{1\to 2}]$ does not depend on the chosen path. We do this by explicit computation using Eq.~(\ref{eq:discrete}), which actually simplifies considerably when taken modulo 2, as the sign of the Jacobian is always equal to $+1\mod 2$.
Let us pick as a reference point for instance the point $\vec n_0=(-1,0,0,\dots)$, in other words, minus the unit vector corresponding to the  positive parity component (let's call this the "south pole" of $S^D$, with $-\vec n_0$ being the "north pole").
Because of $\dhat_{1\to 2} (\k,t)=P\dhat_{1\to 2}(-\k,-t)$, if $(k,t)$ maps to the south pole, so does $(-k,-t)$,  therefore the source points come in pairs unless they correspond to the $D$ dimensional HSPs. 
Again because of parities, the HSPs must map to either the south or north pole.
Then Eq.~(\ref{eq:discrete}) simplifies to
\bea
\label{eq:deg12}
&&\deg_2[\dhat_{1\to 2}]=
\sum_{\substack{(\vec k,t) \\ \dhat_{1\to 2}(\vec k,t)=\vec n_0}}\hspace{-.5 cm}1\mod 2\nn\\
&&=\sum_{\trimk,\bar t}
\frac{1-n^{0}_{1\to 2}(\trimk,\bar t)}{2}\mod 2\label{eq:33}\\
&&=
\sum_{\trimk}\frac{1-n^{0}_{1}(\trimk)}{2}+
\sum_{\trimk}\frac{1-n^{0}_{2}(\trimk)}{2}\mod 2\nn.
\eea
In the second line we have discarded the non-HSPs in the summation (coming in pairs they contribute $0 \mod 2$), and the summand evaluates to one (zero) for HSPs that map to the south pole (north pole), such that the sum can run over all HSPs. The two terms in the last line correspond to $\bar t=0$ and $\bar t=\pi$.
Since the only information that enters on the final expression is contained in $\dhat_1$ and $\dhat_2$, we conclude that 
$\deg_2[\dhat_{1\to 2}]$ is indeed independent of the path chosen and it constitutes a
 well-defined relative invariant of the systems $\dhat_1$ and $\dhat_2$.
It is often more convenient to define the parity invariant
\be
P_{(1)}[\dhat_1,\dhat_2]\equiv
(-1)^{\deg_2[\dhat_{1\to2}]},
\label{eq:34}
\ee
that takes values $+1$ for equivalent and $-1$ for inequivalent Hamiltonians \footnote{The subindex $(1)$ refers to first descendant.} and  is obviously in one to one correspondence with $\deg_2$.

Having defined a relative binary invariant between the systems $\dhat_1$ and $\dhat_2$, we can consider the trivial constant Hamiltonian defined by $\dhat_{\rm ref}(\vec k)=(1,0,0,\dots)$ as a reference and define the absolute $\mathbb Z_2$ invariant
\be
P_{(1)}[\dhat]\equiv 
P_{(1)}[\dhat_{\rm ref},\dhat]=\prod_\trimk n^0(\trimk).
\label{eq:P1}
\ee
where the last equality follows from Eq.~(\ref{eq:33}) and Eq.~(\ref{eq:34}).

We now limit ourselves to the special case with one of the odd components of $\dhat(\k)$ identically zero (taken, without loss of generality $n^{D}$) such that we have exactly $D$ nonzero components 
 and $\vec n(\vec k)$ becomes a map  
\be
 \vec n(\vec k):T^{D-1}\to S^{D-1}.
 \label{eq:firstdes}
\ee 
In this case we can define the quantity $\deg_2[\dhat]$ via Eq.~(\ref{eq:deg}) and compute
\be
\deg_2[\dhat]=\sum_{\trimk}
\frac{1-n^{0}(\trimk)}{2}.
\label{eq:deg2n}
\ee
Comparing with Eq.~(\ref{eq:deg12}) we see that 
\be
\deg_2[\dhat_{1\to 2}]=
\deg_2[\dhat_{1}]+
\deg_2[\dhat_{2}],
\ee
and hence two Hamiltonians are topologically equivalent if and only if
\be
\deg_2[\dhat_{1}]=
\deg_2[\dhat_{2}].
\ee
Then, in this restricted class of Hamiltonians, we have
\be
P_{(1)}[\vec n]=(-1)^{\deg_2[\dhat ]},
\ee
and the topological invariant can be interpreted as wrapping number mod 2 of the map Eq.~(\ref{eq:firstdes}).  This first descendant case covers the Majorana number in 1D class D\cite{Kitaev01}, 2D class DIII realized by TR-invariant TSCs\cite{Qi09,Wang14}, and 3D class AII that most TR-invariant TIs belong to in reality\cite{Fu07_2,Zhang09,Liu10}. 


\subsection{Second descendants}
\label{sec:secondd}

For second descendants in $D-2$ dimensions, we proceed in a similar way. Again, we would like to find a relative invariant between two Hamiltonians defined by $\dhat_1(\k)$ and $\dhat_2(\k)$. Notice that these are now maps from $T^{D-2}\to S^{D}$.
We define a path $\dhat (\k,s)$  with
\be
\dhat_{1\to 2}(\k,0)=\dhat_{1}(\k),\qquad \dhat_{1\to 2}(\k,\pi)=\dhat_{2}(\k),
\ee
 in which $\dhat_{1\to 2}(\k,s)=P\dhat_{1\to 2}(-\k,-s)$. This defines a map $T^{D-1}\to S^{D}$, i.e.~a $D-1$ dimensional Hamiltonian in the same symmetry class which can be thought of as a first descendant. 
As just shown, we can define an invariant $P_{(1)}[\vec n_{1\to 2}]$ for this first descendant.
Provided that this number is independent of the interpolation chosen, we can define
\be
P_{(2)}[\vec n_1, \vec n_2  ]\equiv P_{(1)}[\vec n_{1\to 2}]\,,
\label{eq:P2}
\ee
which constitutes now a relative invariant between the two second descendant Hamiltonians. 
To see that $P_{(1)}[\vec n_{1\to 2}]$ indeed only depends on $\vec n_1$ and $\vec n_2$ and not on the interpolation, we consider a second interpolation
$\vec n'_{1\to2}$ and compute the relative invariant 
$P_{(1)}[\vec n_{1\to 2},\vec n'_{1\to 2}]$. To this end, one defines an "interpolation of interpolations" 
$
\vec n_{12\to 12'}(\k,s,t)
$
that satisfies 
\bea
\vec n_{12\to 12'}(\k,s,0)&=&\vec n_{1\to 2}(\k,s)\nn,\\
\vec n_{12\to 12'}(\k,s,\pi)&=&\vec n'_{1\to 2}(\k,s),
\eea
as well as $n_{12\to 12'}(\k,s,t)=Pn_{12\to 12'}(-\k,-s,-t)$. For this interpolation we can easily compute $\deg_2$.
The computation is just an iteration of Eq.~(\ref{eq:deg12}). Again by using Eq.~(\ref{eq:discrete}) one finds that this is always an even number and hence the two interpolations are equivalent and Eq.~(\ref{eq:P2}) is well defined.

Choosing again the trivial Hamiltonian defined by $\dhat_{\rm ref}=(1,0,0,\dots)$ as a reference, we can define
\be
P_{(2)}[\dhat]\equiv P_{(2)}[\dhat_{\rm ref},\dhat]=\prod_\trimk n^{0}(\trimk).
\label{eq:second}
\ee
Finally, by considering  Hamiltonians with $n^{D-1}=n^{D}=0$, we can identify the right hand side of Eq.~(\ref{eq:second}) with the $D-2$ dimensional wrapping number,
\be
P_{(2)}[\dhat]=(-1)^{\deg[\vec n]}\,,
\ee
of the map $\vec n(\k):T^{D-2}\to S^{D-2}$.  Examples of this second descendant case include 1D class DIII, the ${\mathbb Z}_{2}$ invariant in the quantum spin Hall effect (QSHE) of 2D class AII\cite{Kane05,Kane05_2,Bernevig06,Bernevig06_2,Fu07,Moore07}, and 3D class CII.

\section{Conclusions}

In this paper we have shown that topological invariants for Dirac Hamiltonians  in any dimensions and symmetry classes can be extracted from the wrapping number introduced in Eq.~(\ref{eq:deg}), which describes the degree of the map from the $D$ dimensional BZ $T^D$ to the $D$ dimensional Dirac sphere $S^D$ form by the ${\bf n}({\bf k})$-vector that parametrizes the Dirac Hamiltonian.  Within the context of differential geometry, the wrapping number is simply the number of times the domain $T^D$ covers (or wraps around) the image $S^D$ under this map. By explicitly constructing the Dirac Hamiltonian for each nontrivial symmetry class in any dimension\cite{Ryu10}, we have shown that  all known invariants can be expressed in terms of the wrapping number, as summarized in table \ref{tab:invariants}.
 Moreover, in the case of all binary invariants we have derived a rather simple universal formula that only depend on the values of the Hamiltonian at the HSP points of the BZ, as given by Eq.~(\ref{eq:P12}).

Some of the results in the present paper have been previously obtained in certain special cases. In particular, the concept of the degree of the map has been recognized in Ref.~\onlinecite{Wang_2010} in order to show the equivalence of certain topological invariants. Our work significantly advances this type of approach by combining it with symmetry classification, resulting in a remarkable unification of all known topological invariants, which gives all the physical phenomena related to topological order such as QHE, QAHE, QSHE, charge polarization, Majorana fermions, chiral or helical TSCs, etc, a geometrical interpretation in terms of the Dirac sphere. Moreover, the algebraic formula given by Eq.~(\ref{eq:discrete}) greatly simplifies the practical computation of {\it all} topological invariants, either integer or binary. In the latter case, it also directly leads us to the very simple explicit expression Eq.~(\ref{eq:P12}). Finally, we remark that for the most general Hamiltonians of dimension $2^n$ beyond the Dirac Hamiltonian in Eq.~(\ref{eq:dirac}), one requires the Dirac matrices with $\leq n$ antisymmetric indices  \cite{Bernevig13} to expand the Hamiltonian $H=\sum_{i}r^{i}\Gamma_{i}+\sum_{ij}r^{ij}\Gamma_{ij}+\dots$. Whether the wrapping number in Eq.~(\ref{eq:deg}) can be modified to describe the topology is an intriguing issue that awaits to be clarified. In addition, the cyclic derivative of the ${\bf n}$-vector that enters the integrand of Eq.~(\ref{eq:deg}) can be written in terms of single-particle Green's function for several symmetry classes in low dimensions\cite{Yakovenko90,Hlousek90,Gurarie11,Santos11,Chen18}. Whether such expression holds for all the topologically nontrivial cases on the periodic table remains to be explored.

\begin{table*}
{
\begin{tabular}{ccccccccc}
			&$D=0$		&$D=1$		&$D=2$		&$D=3$	&$D=4$	&$D=5$	&$D=6$	&$D=7$\\
\hline
AI			&$({\cblue A},-)_0$&	&			&		&$({\cblue B},-)_3$&&\la&\la\\
BDI			&\la		&$({\cblue AS},{\cblue A})_1$&&&		&$({\cblue BS},{\cblue B})_4$&&\la\\
D 			&\la		&\la	&$(-,{\cblue A})_1$&&			&		&$(-,{\cblue B})_4$\\
DIII		&			&\la	&\la		&$({\cred A},{\cblue AS})_2$&&&	&$({\cred B},{\cblue BS})_5$\\
AII			&$({\cred B},-)_1$&	&\la		&\la	&$({\cred A},-)_2$\\
CII			&			&$({\cred BS},{\cred B})_2$		&			&\la	&\la&$({\cred AS},{\cred A})_3$\\
C 			&			&		&$(-,{\cred B})_2$			&		&\la&\la&$(-,{\cred A})_3$\\
CI 			&			&		&			&$({\cblue B},{\cred BS})_3$		&	&\la&\la&$({\cblue A},{\cred AS})_4$
\end{tabular}
}
\caption{Pairs $(T,C)_n$ for the primary series (diagonal) and the even series (remaining entries). 
The subindex $n$ indicates the size of the Dirac matrices which are $2^n$ dimensional.
Blue entries satisfy $XX^*=+1$ and red ones $XX^*=-1$, where $X=T,C$.  Here, $B$ is the product of all real $\Gamma_i$, $A=B\Gamma_0$ (where $\Gamma_0$ is the $\Gamma_i$ that plays the role of the "mass term" in the Dirac Lagrangian), and $S$ is the $\Gamma_i$ that implements the chiral symmetry. The arrows indicate the descendants that simply inherit the choices from the primary series.
The table continues periodically for $D>7$, with the index $n$ increasing by 4 with each period.
\label{tab:TC}
}
\end{table*}

\appendix

\section{Dirac Hamiltonians for all nontrivial symmetry classes in general dimension}
\label{sec:details}

The Clifford algebra of $SO(2n+1)$, defined by
\be
\{\Gamma_i,\Gamma_j\}=2\delta_{ij}\,,
\label{eq:CA}
\ee
is generated by $2n+1$ matrices $\Gamma_i$ of dimension $2^n$, which are usually defined recursively.
One particularly simple choice of basis is generated by 
\be
\Gamma_{i}=\Gamma'_i\otimes\sigma_1,\qquad 
\Gamma_{2n}=1\otimes\sigma_2,\qquad
\Gamma_{2n+1}= 1\otimes\sigma_3,
\label{eq:base}
\ee
where $\Gamma'_i$ are the $2n-1$  matrices of the $SO(2n-1)$ Clifford algebra, and we use $\Gamma=1$ for the case $n=0$.
In this convention, the $(n+1)$ odd-index matrices are purely real, and the $n$ even-index matrices are purely imaginary.
The product of all the $2n+1$ matrices is proportional to the identity and in the above basis equal to 
\be
\omega\equiv\Gamma_1\cdots\Gamma_{2n+1}=i^n\,.
\ee
The sign of $\omega$ depends on the choice of basis, but $\omega^2$ is basis independent.
An important quantity is the matrix
\be
B\equiv\Gamma_1\Gamma_3\cdots\Gamma_{2n+1},
\ee
(that is, the product of all real matrices), as it implements complex conjugation on the $\Gamma$ matrices\,,
\be
B\Gamma_iB^{-1}=(-1)^{n} \Gamma_i^*\,,
\ee
which is easily checked using Eq.~(\ref{eq:CA}) and the reality property of the $\Gamma_i$.

We will now build Dirac Hamiltonians for all nontrivial AZ symmetry classes in any dimension $D$ of the form 
\be
H=\sum_{i=0}^D r_i(\vec k)\Gamma_i\,,
\ee
where $\Gamma_0$ is defined explicitly below.
We will follow Ref.~\onlinecite{Ryu10} with slightly different conventions.

For any given $D$, there are one complex and 4 real nontrivial AZ classes.
The complex AZ classes (A and AIII) are simpler and can be constructed from the $n=\floor{\frac{D+1}{2}}$ Clifford algebra and by choosing $\Gamma_0\equiv \Gamma_{D+1}$, where $\floor{...}$ denotes the floor function. The only nontrivial transformation is the chiral transformation $S=\Gamma_{D+2}=\Gamma_{2n+1}$ in the odd-dimensional case.

For the real classes, the functions $\vec r(\vec k)$ have to have definite parities under the reflection of $\vec k$ as a consequence of the $C$ and $T$ transformations in Eq.~(\ref{General_symmetries_TR_PH_CH}).
For definiteness, we will demand that 
\be
r_{0}(\vec k)=r_{0}(-\vec k)\,,\qquad
r_i(\vec k)=-r_i(-\vec k)\,,
\label{eq:parities}
\ee
and construct $C$ and $T$ matrices from this condition.
These are the most relevant cases (for instance for linear Dirac models).
Other choices are possible, see for instance the discussion at the end of section \ref{sec:chern}.

Firstly, choose $n=\floor{\frac{D+1}{2}}$ and
\be
\Gamma_0=\Gamma_{D+1}\,.
\ee
Defining $A\equiv B\Gamma_0$. 
one has
\be
A\Gamma_i^* =(-1)^{n-1}\Gamma_i A,\qquad
A\Gamma_0^*=(-1)^{n}\Gamma_0 A,
\ee
for $1\leq i\leq D$. 
Therefore, choosing $T=A$ ($C=A$) for $n$ even (odd) enforces the parities Eq.~(\ref{eq:parities}) by means of the relations in Eq.~(\ref{General_symmetries_TR_PH_CH}). 
For odd $D$, one has a chiral symmetry with $S\equiv \Gamma_{D+2}=\Gamma_{2n+1}$. 
Moreover, from
\be
AA^*=(-1)^\frac{n(n-1)}{2},\qquad AS(AS)^*=-(-1)^\frac{n(n+1)}{2},
\ee
one infers the AZ classes for the given Hamiltonian. When $D$ changes from 0 to 7 (mod 8), one precisely cycles through  each of the 8 real AZ classes once. This infinite series of Hamiltonians is called the {\it primary series} and is shown in the main diagonal of table \ref{tab:TC}.

Secondly choose $n=\floor{\frac{D+3}{2}}$ and define
\be
\Gamma_0\equiv -i\Gamma_{D+1}\Gamma_{D+2}\Gamma_{D+3}.
\ee
One has then
\be
B\Gamma_i^* =(-1)^{n}\Gamma_i B,\qquad
B\Gamma_0^*=(-1)^{n-1}\Gamma_0 B,
\ee
for $i\leq D$. Thus, for $n$ is even (odd), one can write $T=B$ ($C=B$).
Again, for odd $D$, one can define a chiral symmetry by $S\equiv\Gamma_{D+4}=\Gamma_{2n+1}$, and from 
\be
BB^*=(-1)^\frac{n(n+1)}{2},
\qquad
BS(BS)^*=(-1)^\frac{n(n-1)}{2},
\ee
one infers the AZ symmetry classes. Again, we cycle periodically through all 8 real AZ classes in this series, this is known as the {\it even series} and is shown in the sub-diagonal in table \ref{tab:TC}.

In these conventions, the primary and even series both satisfy
\be
\Gamma_0\Gamma_1\cdots\Gamma_{D}=i^\frac{D}{2},
\label{eq:pseudoeven}
\ee
for even $D$, and 
\be
\Gamma_0\Gamma_1\cdots\Gamma_{D}S=-i^\frac{D+1}{2},
\label{eq:pseudoodd}
\ee
for odd $D$. These relations will be important in the proof of equivalence of Chern/winding number with wrapping number.

The primary and even series cover all real AZ classes with integer topological invariants.
There are two more series, the so-called first and second descendants which have binary topological invariants. These are simply obtained from the primary series by going to one (two) dimensions lower and  setting $r_D=0$ ($r_D=r_{D-1}=0$) respectively. This exhausts all nontrivial real AZ classes in any dimension. 
We close with two comments: (i) The construction of these Hamiltonians is actually basis-independent, for instance one can use a permutation of the $\Gamma_i$ defined in Eq.~(\ref{eq:base}) or any other basis (see the caption of table \ref{tab:TC} for the general expressions for $T$ and $C$ matrices). (ii) the  Hamiltonians of the even series can be made block diagonal, as
\be
[H,\Gamma_j\Gamma_k]=[H,\Gamma_k\Gamma_\ell]=[H,\Gamma_j\Gamma_\ell]=0\,.
\ee 


\section{The degree of the map onto spheres}

\label{sec:deg}

In this appendix we provide some basic information on the degree of the map.
Consider two manifolds, $N$ and $M$ of the same dimension $D$, and a map $f:N\to M$. Let $M$  be orientable with a volume form $\omega$ that is nowhere vanishing, and define
\be
V_M\equiv \int_M\omega\,.
\ee
The integer-valued degree of the map $f$ is defined as
\be
\deg f=\frac{1}{V_M}\int_N f^*\omega,
\ee
where $f^*\omega$ is the pullback of $\omega$  over the map $f$, which is a $D$-form on $N$.
The equivalent algebraic definition of the degree reads 
\be
\deg f=\sum_{k\in f^{-1}(x_0)}\operatorname{sign} J(k), 
\label{eq:algebraic}
\ee
where $J$ is the Jacobian of the map $f$,
\be 
J_f(k)=\det \frac{\partial f^\mu}{\partial k^\nu},
\ee
and $x_0$ is some conveniently chosen base point on $M$ that is non-singular, i.e., $J(k)\neq 0$ for all $k\in f^{-1}(x_0)$.

We now proceed to derive the relevant expressions for the case $M=S^D$ which are used in the main text.
Firstly, when $M$ is Riemannian with metric $g$, we may choose
\be
\omega=\sqrt {\det g(x)}\, d^Dx\,,
\ee
where $x^\mu$ are local coordinates on $M$, and we defined the usual shorthand $d^Dx\equiv dx^1\wedge\dots\wedge dx^D$. 
In terms of local coordinates   $k^\mu$ on $N$, 
\be
f^*\omega=\sqrt{\det g(f(k))}J_f(k)\, d^Dk.
\label{eq:pullback}
\ee
When $M=S^D$, we can embed it in $\mathbb R^{D+1}$ as a unit vector $\vec m(x)^2=1$, and define the canonical metric as 
$g_{\mu\nu}(x)\equiv\frac{\partial\vec m}{\partial x^\mu}\cdot \frac{\partial\vec m}{\partial x^\nu}$.  The volume form can be expressed as
\be
\omega = \frac{1}{D!}\epsilon_{i_0 i_1\cdots i_D} m^{i_0}d m^{i_1}\wedge\dots \wedge m^{i_D},
\ee
that is, in local coordinates
\be
\sqrt{\det g(x)}=
\det \left(\vec m,
	\frac{\partial \vec m}{\partial x^1},\cdots, \frac{\partial \vec m}{\partial x^D}\right).
	\label{eq:detg}
\ee
By defining
\be
\vec n(k)\equiv \vec m(f(k)),
\ee
we can neatly express the pullback of $\omega$ over $f$ 
as
\bea
f^*\omega&=&\frac{1}{D!}\epsilon_{i_0 i_1\cdots i_D} n^{i_0}d n^{i_1}\wedge\dots \wedge n^{i_D}\nn\\
&=&\det \left(\vec n,
	\frac{\partial \vec n}{\partial k^1},\cdots, \frac{\partial \vec n}{\partial k^D}\right)
	d^Dk.
\label{eq:pullbackn}
\eea
Indeed,
since 
\begin{multline}
\det \left(\vec n,
	\frac{\partial \vec n}{\partial k^1},\cdots, \frac{\partial \vec n}{\partial k^D}\right)\\
=J_f(k)\det \left(\vec m,
	\frac{\partial \vec m}{\partial x^1},\cdots, \frac{\partial \vec m}{\partial x^D}\right)_{x=f(k)},
\label{eq:detn}
\end{multline}
Eq.~(\ref{eq:pullbackn}) coincides with Eq.~(\ref{eq:pullback}) by use of Eq.~(\ref{eq:detg}).

Moreover, we note that if $\vec n$ is given as $\vec n=\vec r/|\vec r|$, we also have that 
\be
\det\left(\vec n,
	\frac{\partial \vec n}{\partial k^1},\cdots, \frac{\partial \vec n}{\partial k^D}\right)
=\frac{1}{|\vec r|^{D+1}}
\det \left(\vec r,
	\frac{\partial \vec r}{\partial k^1},\cdots, \frac{\partial \vec r}{\partial k^D}\right)
	\label{eq:nr},
\ee
which allows us to exchange $\vec n$ for $\vec r$ in many formulas.
Finally, since $\sqrt{\det g}>0$ we also have from Eq.~(\ref{eq:detn}) and (\ref{eq:nr})
\bea
\operatorname{sign} J_f(k) &=& \operatorname{sign}\det
\left(\vec n,
	\frac{\partial \vec n}{\partial k^1},\cdots, \frac{\partial \vec n}{\partial k^D}\right)\nn\\
&=&\operatorname{sign}\det
\left(\vec r,
	\frac{\partial \vec r}{\partial k^1},\cdots, \frac{\partial \vec r}{\partial k^D}\right),	
\eea
which simplifies the calculation of Eq.~(\ref{eq:algebraic}).

\bibliography{Literatur}

\begin{thebibliography}{60}%
\makeatletter
\providecommand \@ifxundefined [1]{%
 \@ifx{#1\undefined}
}%
\providecommand \@ifnum [1]{%
 \ifnum #1\expandafter \@firstoftwo
 \else \expandafter \@secondoftwo
 \fi
}%
\providecommand \@ifx [1]{%
 \ifx #1\expandafter \@firstoftwo
 \else \expandafter \@secondoftwo
 \fi
}%
\providecommand \natexlab [1]{#1}%
\providecommand \enquote  [1]{``#1''}%
\providecommand \bibnamefont  [1]{#1}%
\providecommand \bibfnamefont [1]{#1}%
\providecommand \citenamefont [1]{#1}%
\providecommand \href@noop [0]{\@secondoftwo}%
\providecommand \href [0]{\begingroup \@sanitize@url \@href}%
\providecommand \@href[1]{\@@startlink{#1}\@@href}%
\providecommand \@@href[1]{\endgroup#1\@@endlink}%
\providecommand \@sanitize@url [0]{\catcode `\\12\catcode `\$12\catcode
  `\&12\catcode `\#12\catcode `\^12\catcode `\_12\catcode `\%12\relax}%
\providecommand \@@startlink[1]{}%
\providecommand \@@endlink[0]{}%
\providecommand \url  [0]{\begingroup\@sanitize@url \@url }%
\providecommand \@url [1]{\endgroup\@href {#1}{\urlprefix }}%
\providecommand \urlprefix  [0]{URL }%
\providecommand \Eprint [0]{\href }%
\providecommand \doibase [0]{http://dx.doi.org/}%
\providecommand \selectlanguage [0]{\@gobble}%
\providecommand \bibinfo  [0]{\@secondoftwo}%
\providecommand \bibfield  [0]{\@secondoftwo}%
\providecommand \translation [1]{[#1]}%
\providecommand \BibitemOpen [0]{}%
\providecommand \bibitemStop [0]{}%
\providecommand \bibitemNoStop [0]{.\EOS\space}%
\providecommand \EOS [0]{\spacefactor3000\relax}%
\providecommand \BibitemShut  [1]{\csname bibitem#1\endcsname}%
\let\auto@bib@innerbib\@empty
\bibitem [{\citenamefont {Schnyder}\ \emph {et~al.}(2008)\citenamefont
  {Schnyder}, \citenamefont {Ryu}, \citenamefont {Furusaki},\ and\
  \citenamefont {Ludwig}}]{Schnyder08}%
  \BibitemOpen
  \bibfield  {author} {\bibinfo {author} {\bibfnamefont {A.~P.}\ \bibnamefont
  {Schnyder}}, \bibinfo {author} {\bibfnamefont {S.}~\bibnamefont {Ryu}},
  \bibinfo {author} {\bibfnamefont {A.}~\bibnamefont {Furusaki}}, \ and\
  \bibinfo {author} {\bibfnamefont {A.~W.~W.}\ \bibnamefont {Ludwig}},\ }\href
  {\doibase 10.1103/PhysRevB.78.195125} {\bibfield  {journal} {\bibinfo
  {journal} {Phys. Rev. B}\ }\textbf {\bibinfo {volume} {78}},\ \bibinfo
  {pages} {195125} (\bibinfo {year} {2008})}\BibitemShut {NoStop}%
\bibitem [{\citenamefont {Ryu}\ \emph {et~al.}(2010)\citenamefont {Ryu},
  \citenamefont {Schnyder}, \citenamefont {Furusaki},\ and\ \citenamefont
  {Ludwig}}]{Ryu10}%
  \BibitemOpen
  \bibfield  {author} {\bibinfo {author} {\bibfnamefont {S.}~\bibnamefont
  {Ryu}}, \bibinfo {author} {\bibfnamefont {A.~P.}\ \bibnamefont {Schnyder}},
  \bibinfo {author} {\bibfnamefont {A.}~\bibnamefont {Furusaki}}, \ and\
  \bibinfo {author} {\bibfnamefont {A.~W.~W.}\ \bibnamefont {Ludwig}},\ }\href
  {http://stacks.iop.org/1367-2630/12/i=6/a=065010} {\bibfield  {journal}
  {\bibinfo  {journal} {New J. Phys.}\ }\textbf {\bibinfo {volume} {12}},\
  \bibinfo {pages} {065010} (\bibinfo {year} {2010})}\BibitemShut {NoStop}%
\bibitem [{\citenamefont {Kitaev}(2009)}]{Kitaev09}%
  \BibitemOpen
  \bibfield  {author} {\bibinfo {author} {\bibfnamefont {A.}~\bibnamefont
  {Kitaev}},\ }\href {\doibase 10.1063/1.3149495} {\bibfield  {journal}
  {\bibinfo  {journal} {AIP Conf. Proc.}\ }\textbf {\bibinfo {volume} {1134}},\
  \bibinfo {pages} {22} (\bibinfo {year} {2009})}\BibitemShut {NoStop}%
\bibitem [{\citenamefont {Chiu}\ \emph {et~al.}(2016)\citenamefont {Chiu},
  \citenamefont {Teo}, \citenamefont {Schnyder},\ and\ \citenamefont
  {Ryu}}]{Chiu16}%
  \BibitemOpen
  \bibfield  {author} {\bibinfo {author} {\bibfnamefont {C.-K.}\ \bibnamefont
  {Chiu}}, \bibinfo {author} {\bibfnamefont {J.~C.~Y.}\ \bibnamefont {Teo}},
  \bibinfo {author} {\bibfnamefont {A.~P.}\ \bibnamefont {Schnyder}}, \ and\
  \bibinfo {author} {\bibfnamefont {S.}~\bibnamefont {Ryu}},\ }\href {\doibase
  10.1103/RevModPhys.88.035005} {\bibfield  {journal} {\bibinfo  {journal}
  {Rev. Mod. Phys.}\ }\textbf {\bibinfo {volume} {88}},\ \bibinfo {pages}
  {035005} (\bibinfo {year} {2016})}\BibitemShut {NoStop}%
\bibitem [{\citenamefont {Qi}\ \emph {et~al.}(2008)\citenamefont {Qi},
  \citenamefont {Hughes},\ and\ \citenamefont {Zhang}}]{Qi08}%
  \BibitemOpen
  \bibfield  {author} {\bibinfo {author} {\bibfnamefont {X.-L.}\ \bibnamefont
  {Qi}}, \bibinfo {author} {\bibfnamefont {T.~L.}\ \bibnamefont {Hughes}}, \
  and\ \bibinfo {author} {\bibfnamefont {S.-C.}\ \bibnamefont {Zhang}},\ }\href
  {\doibase 10.1103/PhysRevB.78.195424} {\bibfield  {journal} {\bibinfo
  {journal} {Phys. Rev. B}\ }\textbf {\bibinfo {volume} {78}},\ \bibinfo
  {pages} {195424} (\bibinfo {year} {2008})}\BibitemShut {NoStop}%
\bibitem [{\citenamefont {{Kronecker}}(1869)}]{Kronecker1869}%
  \BibitemOpen
  \bibfield  {author} {\bibinfo {author} {\bibfnamefont {L.}~\bibnamefont
  {{Kronecker}}},\ }\href@noop {} {\bibfield  {journal} {\bibinfo  {journal}
  {{Berl. Monatsber.}}\ }\textbf {\bibinfo {volume} {1869}},\ \bibinfo {pages}
  {159} (\bibinfo {year} {1869})}\BibitemShut {NoStop}%
\bibitem [{\citenamefont {{Brouwer}}(1912)}]{Brouwer1912}%
  \BibitemOpen
  \bibfield  {author} {\bibinfo {author} {\bibfnamefont {L.}~\bibnamefont
  {{Brouwer}}},\ }\href@noop {} {\bibfield  {journal} {\bibinfo  {journal}
  {{Math. Ann.}}\ }\textbf {\bibinfo {volume} {71}},\ \bibinfo {pages} {97}
  (\bibinfo {year} {1912})}\BibitemShut {NoStop}%
\bibitem [{\citenamefont {Altland}\ and\ \citenamefont
  {Zirnbauer}(1997)}]{Altland97}%
  \BibitemOpen
  \bibfield  {author} {\bibinfo {author} {\bibfnamefont {A.}~\bibnamefont
  {Altland}}\ and\ \bibinfo {author} {\bibfnamefont {M.~R.}\ \bibnamefont
  {Zirnbauer}},\ }\href {\doibase 10.1103/PhysRevB.55.1142} {\bibfield
  {journal} {\bibinfo  {journal} {Phys. Rev. B}\ }\textbf {\bibinfo {volume}
  {55}},\ \bibinfo {pages} {1142} (\bibinfo {year} {1997})}\BibitemShut
  {NoStop}%
\bibitem [{\citenamefont {Chen}\ \emph {et~al.}(2017)\citenamefont {Chen},
  \citenamefont {Legner}, \citenamefont {R\"uegg},\ and\ \citenamefont
  {Sigrist}}]{Chen17}%
  \BibitemOpen
  \bibfield  {author} {\bibinfo {author} {\bibfnamefont {W.}~\bibnamefont
  {Chen}}, \bibinfo {author} {\bibfnamefont {M.}~\bibnamefont {Legner}},
  \bibinfo {author} {\bibfnamefont {A.}~\bibnamefont {R\"uegg}}, \ and\
  \bibinfo {author} {\bibfnamefont {M.}~\bibnamefont {Sigrist}},\ }\href
  {\doibase 10.1103/PhysRevB.95.075116} {\bibfield  {journal} {\bibinfo
  {journal} {Phys. Rev. B}\ }\textbf {\bibinfo {volume} {95}},\ \bibinfo
  {pages} {075116} (\bibinfo {year} {2017})}\BibitemShut {NoStop}%
\bibitem [{\citenamefont {Chen}\ and\ \citenamefont
  {Sigrist}(2019)}]{Chen19_AMS_review}%
  \BibitemOpen
  \bibfield  {author} {\bibinfo {author} {\bibfnamefont {W.}~\bibnamefont
  {Chen}}\ and\ \bibinfo {author} {\bibfnamefont {M.}~\bibnamefont {Sigrist}},\
  }\href@noop {} {\emph {\bibinfo {title} {Advanced Topological Insulators, Ch.
  7}}}\ (\bibinfo  {publisher} {Wiley-Scrivener},\ \bibinfo {year}
  {2019})\BibitemShut {NoStop}%
\bibitem [{\citenamefont {Molignini}\ \emph {et~al.}(2018)\citenamefont
  {Molignini}, \citenamefont {Chen},\ and\ \citenamefont
  {Chitra}}]{Molignini18_Floquet_Majorana}%
  \BibitemOpen
  \bibfield  {author} {\bibinfo {author} {\bibfnamefont {P.}~\bibnamefont
  {Molignini}}, \bibinfo {author} {\bibfnamefont {W.}~\bibnamefont {Chen}}, \
  and\ \bibinfo {author} {\bibfnamefont {R.}~\bibnamefont {Chitra}},\ }\href
  {\doibase 10.1103/PhysRevB.98.125129} {\bibfield  {journal} {\bibinfo
  {journal} {Phys. Rev. B}\ }\textbf {\bibinfo {volume} {98}},\ \bibinfo
  {pages} {125129} (\bibinfo {year} {2018})}\BibitemShut {NoStop}%
\bibitem [{\citenamefont {Molignini}\ \emph {et~al.}(2020)\citenamefont
  {Molignini}, \citenamefont {Chen},\ and\ \citenamefont
  {Chitra}}]{Molignini20_multicritical}%
  \BibitemOpen
  \bibfield  {author} {\bibinfo {author} {\bibfnamefont {P.}~\bibnamefont
  {Molignini}}, \bibinfo {author} {\bibfnamefont {W.}~\bibnamefont {Chen}}, \
  and\ \bibinfo {author} {\bibfnamefont {R.}~\bibnamefont {Chitra}},\ }\href
  {\doibase 10.1103/PhysRevB.101.165106} {\bibfield  {journal} {\bibinfo
  {journal} {Phys. Rev. B}\ }\textbf {\bibinfo {volume} {101}},\ \bibinfo
  {pages} {165106} (\bibinfo {year} {2020})}\BibitemShut {NoStop}%
\bibitem [{\citenamefont {Rufo}\ \emph {et~al.}(2019)\citenamefont {Rufo},
  \citenamefont {Lopes}, \citenamefont {Continentino},\ and\ \citenamefont
  {Griffith}}]{Rufo19}%
  \BibitemOpen
  \bibfield  {author} {\bibinfo {author} {\bibfnamefont {S.}~\bibnamefont
  {Rufo}}, \bibinfo {author} {\bibfnamefont {N.}~\bibnamefont {Lopes}},
  \bibinfo {author} {\bibfnamefont {M.~A.}\ \bibnamefont {Continentino}}, \
  and\ \bibinfo {author} {\bibfnamefont {M.~A.~R.}\ \bibnamefont {Griffith}},\
  }\href {\doibase 10.1103/PhysRevB.100.195432} {\bibfield  {journal} {\bibinfo
   {journal} {Phys. Rev. B}\ }\textbf {\bibinfo {volume} {100}},\ \bibinfo
  {pages} {195432} (\bibinfo {year} {2019})}\BibitemShut {NoStop}%
\bibitem [{\citenamefont {Abdulla}\ \emph {et~al.}(2020)\citenamefont
  {Abdulla}, \citenamefont {Mohan},\ and\ \citenamefont {Rao}}]{Abdulla20}%
  \BibitemOpen
  \bibfield  {author} {\bibinfo {author} {\bibfnamefont {F.}~\bibnamefont
  {Abdulla}}, \bibinfo {author} {\bibfnamefont {P.}~\bibnamefont {Mohan}}, \
  and\ \bibinfo {author} {\bibfnamefont {S.}~\bibnamefont {Rao}},\ }\href
  {\doibase 10.1103/PhysRevB.102.235129} {\bibfield  {journal} {\bibinfo
  {journal} {Phys. Rev. B}\ }\textbf {\bibinfo {volume} {102}},\ \bibinfo
  {pages} {235129} (\bibinfo {year} {2020})}\BibitemShut {NoStop}%
\bibitem [{\citenamefont {Kumar}\ \emph {et~al.}(2020)\citenamefont {Kumar},
  \citenamefont {Kartik}, \citenamefont {Rahul},\ and\ \citenamefont
  {Sarkar}}]{Kumar20}%
  \BibitemOpen
  \bibfield  {author} {\bibinfo {author} {\bibfnamefont {R.~R.}\ \bibnamefont
  {Kumar}}, \bibinfo {author} {\bibfnamefont {Y.~R.}\ \bibnamefont {Kartik}},
  \bibinfo {author} {\bibfnamefont {S.}~\bibnamefont {Rahul}}, \ and\ \bibinfo
  {author} {\bibfnamefont {S.}~\bibnamefont {Sarkar}},\ }\href {\doibase
  10.1038/s41598-020-80337-7} {\bibfield  {journal} {\bibinfo  {journal} {Sci.
  Rep.}\ }\textbf {\bibinfo {volume} {11}},\ \bibinfo {pages} {1004} (\bibinfo
  {year} {2020})}\BibitemShut {NoStop}%
\bibitem [{\citenamefont {Chen}\ and\ \citenamefont {Schnyder}(2019)}]{Chen19}%
  \BibitemOpen
  \bibfield  {author} {\bibinfo {author} {\bibfnamefont {W.}~\bibnamefont
  {Chen}}\ and\ \bibinfo {author} {\bibfnamefont {A.~P.}\ \bibnamefont
  {Schnyder}},\ }\href {\doibase 10.1088/1367-2630/ab2a2d} {\bibfield
  {journal} {\bibinfo  {journal} {New J. Phys.}\ }\textbf {\bibinfo {volume}
  {21}},\ \bibinfo {pages} {073003} (\bibinfo {year} {2019})}\BibitemShut
  {NoStop}%
\bibitem [{\citenamefont {Panahiyan}\ \emph {et~al.}(2020)\citenamefont
  {Panahiyan}, \citenamefont {Chen},\ and\ \citenamefont
  {Fritzsche}}]{Panahiyan20_fidelity_susceptibility}%
  \BibitemOpen
  \bibfield  {author} {\bibinfo {author} {\bibfnamefont {S.}~\bibnamefont
  {Panahiyan}}, \bibinfo {author} {\bibfnamefont {W.}~\bibnamefont {Chen}}, \
  and\ \bibinfo {author} {\bibfnamefont {S.}~\bibnamefont {Fritzsche}},\ }\href
  {\doibase 10.1103/PhysRevB.102.134111} {\bibfield  {journal} {\bibinfo
  {journal} {Phys. Rev. B}\ }\textbf {\bibinfo {volume} {102}},\ \bibinfo
  {pages} {134111} (\bibinfo {year} {2020})}\BibitemShut {NoStop}%
\bibitem [{\citenamefont {Molignini}\ \emph {et~al.}()\citenamefont
  {Molignini}, \citenamefont {Celades}, \citenamefont {Chitra},\ and\
  \citenamefont {Chen}}]{Molignini21_Kitaev_cross_dim}%
  \BibitemOpen
  \bibfield  {author} {\bibinfo {author} {\bibfnamefont {P.}~\bibnamefont
  {Molignini}}, \bibinfo {author} {\bibfnamefont {A.~G.}\ \bibnamefont
  {Celades}}, \bibinfo {author} {\bibfnamefont {R.}~\bibnamefont {Chitra}}, \
  and\ \bibinfo {author} {\bibfnamefont {W.}~\bibnamefont {Chen}},\ }\href@noop
  {} {\ ,\ \bibinfo {pages} {arXiv:2102.00009}}\BibitemShut {NoStop}%
\bibitem [{\citenamefont {Chen}(2016)}]{Chen16}%
  \BibitemOpen
  \bibfield  {author} {\bibinfo {author} {\bibfnamefont {W.}~\bibnamefont
  {Chen}},\ }\href {http://stacks.iop.org/0953-8984/28/i=5/a=055601} {\bibfield
   {journal} {\bibinfo  {journal} {J. Phys. Condens. Matter}\ }\textbf
  {\bibinfo {volume} {28}},\ \bibinfo {pages} {055601} (\bibinfo {year}
  {2016})}\BibitemShut {NoStop}%
\bibitem [{\citenamefont {Chen}\ \emph {et~al.}(2016)\citenamefont {Chen},
  \citenamefont {Sigrist},\ and\ \citenamefont {Schnyder}}]{Chen16_2}%
  \BibitemOpen
  \bibfield  {author} {\bibinfo {author} {\bibfnamefont {W.}~\bibnamefont
  {Chen}}, \bibinfo {author} {\bibfnamefont {M.}~\bibnamefont {Sigrist}}, \
  and\ \bibinfo {author} {\bibfnamefont {A.~P.}\ \bibnamefont {Schnyder}},\
  }\href {http://stacks.iop.org/0953-8984/28/i=36/a=365501} {\bibfield
  {journal} {\bibinfo  {journal} {J. Phys. Condens. Matter}\ }\textbf {\bibinfo
  {volume} {28}},\ \bibinfo {pages} {365501} (\bibinfo {year}
  {2016})}\BibitemShut {NoStop}%
\bibitem [{\citenamefont {Su}\ \emph {et~al.}(1979)\citenamefont {Su},
  \citenamefont {Schrieffer},\ and\ \citenamefont {Heeger}}]{Su79}%
  \BibitemOpen
  \bibfield  {author} {\bibinfo {author} {\bibfnamefont {W.~P.}\ \bibnamefont
  {Su}}, \bibinfo {author} {\bibfnamefont {J.~R.}\ \bibnamefont {Schrieffer}},
  \ and\ \bibinfo {author} {\bibfnamefont {A.~J.}\ \bibnamefont {Heeger}},\
  }\href {\doibase 10.1103/PhysRevLett.42.1698} {\bibfield  {journal} {\bibinfo
   {journal} {Phys. Rev. Lett.}\ }\textbf {\bibinfo {volume} {42}},\ \bibinfo
  {pages} {1698} (\bibinfo {year} {1979})}\BibitemShut {NoStop}%
\bibitem [{\citenamefont {Bernevig}\ and\ \citenamefont
  {Hughes}(2013)}]{Bernevig13}%
  \BibitemOpen
  \bibfield  {author} {\bibinfo {author} {\bibfnamefont {B.~A.}\ \bibnamefont
  {Bernevig}}\ and\ \bibinfo {author} {\bibfnamefont {T.~L.}\ \bibnamefont
  {Hughes}},\ }\href@noop {} {\emph {\bibinfo {title} {Topological Insulators
  and Topological Superconductors}}}\ (\bibinfo  {publisher} {Princeton
  University Press},\ \bibinfo {year} {2013})\BibitemShut {NoStop}%
\bibitem [{Note1()}]{Note1}%
  \BibitemOpen
  \bibinfo {note} {In a basis where $ S=\Gamma _{2n+1}$ the Hamiltonian is off
  block diagonal and the winding number can be written in terms of the
  off-diagonal block matrices of dimension $N/2$.}\BibitemShut {Stop}%
\bibitem [{\citenamefont {King-Smith}\ and\ \citenamefont
  {Vanderbilt}(1993)}]{KingSmith93}%
  \BibitemOpen
  \bibfield  {author} {\bibinfo {author} {\bibfnamefont {R.~D.}\ \bibnamefont
  {King-Smith}}\ and\ \bibinfo {author} {\bibfnamefont {D.}~\bibnamefont
  {Vanderbilt}},\ }\href {\doibase 10.1103/PhysRevB.47.1651} {\bibfield
  {journal} {\bibinfo  {journal} {Phys. Rev. B}\ }\textbf {\bibinfo {volume}
  {47}},\ \bibinfo {pages} {1651} (\bibinfo {year} {1993})}\BibitemShut
  {NoStop}%
\bibitem [{\citenamefont {Resta}(1994)}]{Resta94}%
  \BibitemOpen
  \bibfield  {author} {\bibinfo {author} {\bibfnamefont {R.}~\bibnamefont
  {Resta}},\ }\href {\doibase 10.1103/RevModPhys.66.899} {\bibfield  {journal}
  {\bibinfo  {journal} {Rev. Mod. Phys.}\ }\textbf {\bibinfo {volume} {66}},\
  \bibinfo {pages} {899} (\bibinfo {year} {1994})}\BibitemShut {NoStop}%
\bibitem [{\citenamefont {Fu}\ and\ \citenamefont {Berg}(2010)}]{Fu10}%
  \BibitemOpen
  \bibfield  {author} {\bibinfo {author} {\bibfnamefont {L.}~\bibnamefont
  {Fu}}\ and\ \bibinfo {author} {\bibfnamefont {E.}~\bibnamefont {Berg}},\
  }\href {\doibase 10.1103/PhysRevLett.105.097001} {\bibfield  {journal}
  {\bibinfo  {journal} {Phys. Rev. Lett.}\ }\textbf {\bibinfo {volume} {105}},\
  \bibinfo {pages} {097001} (\bibinfo {year} {2010})}\BibitemShut {NoStop}%
\bibitem [{\citenamefont {Zhao}\ and\ \citenamefont {Wang}(2014)}]{Zhao14}%
  \BibitemOpen
  \bibfield  {author} {\bibinfo {author} {\bibfnamefont {Y.~X.}\ \bibnamefont
  {Zhao}}\ and\ \bibinfo {author} {\bibfnamefont {Z.~D.}\ \bibnamefont
  {Wang}},\ }\href {\doibase 10.1103/PhysRevB.90.115158} {\bibfield  {journal}
  {\bibinfo  {journal} {Phys. Rev. B}\ }\textbf {\bibinfo {volume} {90}},\
  \bibinfo {pages} {115158} (\bibinfo {year} {2014})}\BibitemShut {NoStop}%
\bibitem [{\citenamefont {Malard}\ \emph {et~al.}(2020)\citenamefont {Malard},
  \citenamefont {Johannesson},\ and\ \citenamefont
  {Chen}}]{Malard20_multicritical}%
  \BibitemOpen
  \bibfield  {author} {\bibinfo {author} {\bibfnamefont {M.}~\bibnamefont
  {Malard}}, \bibinfo {author} {\bibfnamefont {H.}~\bibnamefont {Johannesson}},
  \ and\ \bibinfo {author} {\bibfnamefont {W.}~\bibnamefont {Chen}},\ }\href
  {\doibase 10.1103/PhysRevB.102.205420} {\bibfield  {journal} {\bibinfo
  {journal} {Phys. Rev. B}\ }\textbf {\bibinfo {volume} {102}},\ \bibinfo
  {pages} {205420} (\bibinfo {year} {2020})}\BibitemShut {NoStop}%
\bibitem [{\citenamefont {Thouless}\ \emph {et~al.}(1982)\citenamefont
  {Thouless}, \citenamefont {Kohmoto}, \citenamefont {Nightingale},\ and\
  \citenamefont {den Nijs}}]{Thouless82}%
  \BibitemOpen
  \bibfield  {author} {\bibinfo {author} {\bibfnamefont {D.~J.}\ \bibnamefont
  {Thouless}}, \bibinfo {author} {\bibfnamefont {M.}~\bibnamefont {Kohmoto}},
  \bibinfo {author} {\bibfnamefont {M.~P.}\ \bibnamefont {Nightingale}}, \ and\
  \bibinfo {author} {\bibfnamefont {M.}~\bibnamefont {den Nijs}},\ }\href
  {\doibase 10.1103/PhysRevLett.49.405} {\bibfield  {journal} {\bibinfo
  {journal} {Phys. Rev. Lett.}\ }\textbf {\bibinfo {volume} {49}},\ \bibinfo
  {pages} {405} (\bibinfo {year} {1982})}\BibitemShut {NoStop}%
\bibitem [{\citenamefont {Niu}\ \emph {et~al.}(1985)\citenamefont {Niu},
  \citenamefont {Thouless},\ and\ \citenamefont {Wu}}]{Niu85}%
  \BibitemOpen
  \bibfield  {author} {\bibinfo {author} {\bibfnamefont {Q.}~\bibnamefont
  {Niu}}, \bibinfo {author} {\bibfnamefont {D.~J.}\ \bibnamefont {Thouless}}, \
  and\ \bibinfo {author} {\bibfnamefont {Y.-S.}\ \bibnamefont {Wu}},\ }\href
  {\doibase 10.1103/PhysRevB.31.3372} {\bibfield  {journal} {\bibinfo
  {journal} {Phys. Rev. B}\ }\textbf {\bibinfo {volume} {31}},\ \bibinfo
  {pages} {3372} (\bibinfo {year} {1985})}\BibitemShut {NoStop}%
\bibitem [{\citenamefont {Haldane}(1988)}]{Haldane88}%
  \BibitemOpen
  \bibfield  {author} {\bibinfo {author} {\bibfnamefont {F.~D.~M.}\
  \bibnamefont {Haldane}},\ }\href {\doibase 10.1103/PhysRevLett.61.2015}
  {\bibfield  {journal} {\bibinfo  {journal} {Phys. Rev. Lett.}\ }\textbf
  {\bibinfo {volume} {61}},\ \bibinfo {pages} {2015} (\bibinfo {year}
  {1988})}\BibitemShut {NoStop}%
\bibitem [{\citenamefont {Liu}\ \emph {et~al.}(2016)\citenamefont {Liu},
  \citenamefont {Zhang},\ and\ \citenamefont {Qi}}]{Liu16}%
  \BibitemOpen
  \bibfield  {author} {\bibinfo {author} {\bibfnamefont {C.-X.}\ \bibnamefont
  {Liu}}, \bibinfo {author} {\bibfnamefont {S.-C.}\ \bibnamefont {Zhang}}, \
  and\ \bibinfo {author} {\bibfnamefont {X.-L.}\ \bibnamefont {Qi}},\ }\href
  {\doibase 10.1146/annurev-conmatphys-031115-011417} {\bibfield  {journal}
  {\bibinfo  {journal} {Annu. Rev. Condens. Matter Phys.}\ }\textbf {\bibinfo
  {volume} {7}},\ \bibinfo {pages} {301} (\bibinfo {year} {2016})}\BibitemShut
  {NoStop}%
\bibitem [{\citenamefont {Xiao}\ \emph {et~al.}(2005)\citenamefont {Xiao},
  \citenamefont {Shi},\ and\ \citenamefont {Niu}}]{Xiao05}%
  \BibitemOpen
  \bibfield  {author} {\bibinfo {author} {\bibfnamefont {D.}~\bibnamefont
  {Xiao}}, \bibinfo {author} {\bibfnamefont {J.}~\bibnamefont {Shi}}, \ and\
  \bibinfo {author} {\bibfnamefont {Q.}~\bibnamefont {Niu}},\ }\href {\doibase
  10.1103/PhysRevLett.95.137204} {\bibfield  {journal} {\bibinfo  {journal}
  {Phys. Rev. Lett.}\ }\textbf {\bibinfo {volume} {95}},\ \bibinfo {pages}
  {137204} (\bibinfo {year} {2005})}\BibitemShut {NoStop}%
\bibitem [{\citenamefont {Xiao}\ \emph {et~al.}(2010)\citenamefont {Xiao},
  \citenamefont {Chang},\ and\ \citenamefont {Niu}}]{Xiao10}%
  \BibitemOpen
  \bibfield  {author} {\bibinfo {author} {\bibfnamefont {D.}~\bibnamefont
  {Xiao}}, \bibinfo {author} {\bibfnamefont {M.-C.}\ \bibnamefont {Chang}}, \
  and\ \bibinfo {author} {\bibfnamefont {Q.}~\bibnamefont {Niu}},\ }\href
  {\doibase 10.1103/RevModPhys.82.1959} {\bibfield  {journal} {\bibinfo
  {journal} {Rev. Mod. Phys.}\ }\textbf {\bibinfo {volume} {82}},\ \bibinfo
  {pages} {1959} (\bibinfo {year} {2010})}\BibitemShut {NoStop}%
\bibitem [{\citenamefont {Thonhauser}\ \emph {et~al.}(2005)\citenamefont
  {Thonhauser}, \citenamefont {Ceresoli}, \citenamefont {Vanderbilt},\ and\
  \citenamefont {Resta}}]{Thonhauser05}%
  \BibitemOpen
  \bibfield  {author} {\bibinfo {author} {\bibfnamefont {T.}~\bibnamefont
  {Thonhauser}}, \bibinfo {author} {\bibfnamefont {D.}~\bibnamefont
  {Ceresoli}}, \bibinfo {author} {\bibfnamefont {D.}~\bibnamefont
  {Vanderbilt}}, \ and\ \bibinfo {author} {\bibfnamefont {R.}~\bibnamefont
  {Resta}},\ }\href {\doibase 10.1103/PhysRevLett.95.137205} {\bibfield
  {journal} {\bibinfo  {journal} {Phys. Rev. Lett.}\ }\textbf {\bibinfo
  {volume} {95}},\ \bibinfo {pages} {137205} (\bibinfo {year}
  {2005})}\BibitemShut {NoStop}%
\bibitem [{\citenamefont {Ceresoli}\ \emph {et~al.}(2006)\citenamefont
  {Ceresoli}, \citenamefont {Thonhauser}, \citenamefont {Vanderbilt},\ and\
  \citenamefont {Resta}}]{Ceresoli06}%
  \BibitemOpen
  \bibfield  {author} {\bibinfo {author} {\bibfnamefont {D.}~\bibnamefont
  {Ceresoli}}, \bibinfo {author} {\bibfnamefont {T.}~\bibnamefont
  {Thonhauser}}, \bibinfo {author} {\bibfnamefont {D.}~\bibnamefont
  {Vanderbilt}}, \ and\ \bibinfo {author} {\bibfnamefont {R.}~\bibnamefont
  {Resta}},\ }\href {\doibase 10.1103/PhysRevB.74.024408} {\bibfield  {journal}
  {\bibinfo  {journal} {Phys. Rev. B}\ }\textbf {\bibinfo {volume} {74}},\
  \bibinfo {pages} {024408} (\bibinfo {year} {2006})}\BibitemShut {NoStop}%
\bibitem [{\citenamefont {Shi}\ \emph {et~al.}(2007)\citenamefont {Shi},
  \citenamefont {Vignale}, \citenamefont {Xiao},\ and\ \citenamefont
  {Niu}}]{Shi07}%
  \BibitemOpen
  \bibfield  {author} {\bibinfo {author} {\bibfnamefont {J.}~\bibnamefont
  {Shi}}, \bibinfo {author} {\bibfnamefont {G.}~\bibnamefont {Vignale}},
  \bibinfo {author} {\bibfnamefont {D.}~\bibnamefont {Xiao}}, \ and\ \bibinfo
  {author} {\bibfnamefont {Q.}~\bibnamefont {Niu}},\ }\href {\doibase
  10.1103/PhysRevLett.99.197202} {\bibfield  {journal} {\bibinfo  {journal}
  {Phys. Rev. Lett.}\ }\textbf {\bibinfo {volume} {99}},\ \bibinfo {pages}
  {197202} (\bibinfo {year} {2007})}\BibitemShut {NoStop}%
\bibitem [{\citenamefont {Souza}\ and\ \citenamefont
  {Vanderbilt}(2008)}]{Souza08}%
  \BibitemOpen
  \bibfield  {author} {\bibinfo {author} {\bibfnamefont {I.}~\bibnamefont
  {Souza}}\ and\ \bibinfo {author} {\bibfnamefont {D.}~\bibnamefont
  {Vanderbilt}},\ }\href {\doibase 10.1103/PhysRevB.77.054438} {\bibfield
  {journal} {\bibinfo  {journal} {Phys. Rev. B}\ }\textbf {\bibinfo {volume}
  {77}},\ \bibinfo {pages} {054438} (\bibinfo {year} {2008})}\BibitemShut
  {NoStop}%
\bibitem [{\citenamefont {Read}\ and\ \citenamefont {Green}(2000)}]{Read00}%
  \BibitemOpen
  \bibfield  {author} {\bibinfo {author} {\bibfnamefont {N.}~\bibnamefont
  {Read}}\ and\ \bibinfo {author} {\bibfnamefont {D.}~\bibnamefont {Green}},\
  }\href {\doibase 10.1103/PhysRevB.61.10267} {\bibfield  {journal} {\bibinfo
  {journal} {Phys. Rev. B}\ }\textbf {\bibinfo {volume} {61}},\ \bibinfo
  {pages} {10267} (\bibinfo {year} {2000})}\BibitemShut {NoStop}%
\bibitem [{Note2()}]{Note2}%
  \BibitemOpen
  \bibinfo {note} {However, the $C$ and $T$ symmetries do not enforce this and
  the generic Dirac Hamiltonian of the given symmetry class is defined by
  Eq.~(\ref {eq:first}) and Eq.~(\ref {eq:second}) together with Eq.~(\ref
  {eq:paritiesdes}).}\BibitemShut {Stop}%
\bibitem [{Note3()}]{Note3}%
  \BibitemOpen
  \bibinfo {note} {It is maybe worth pointing out that for fixed times $t\not
  =0,\pi $ the system does not belong to the $D-1$ dimensional symmetry class,
  so the interpolation is not to be regarded as a homotopy within the space of
  $D-1$ dimensional Hamiltonians of that class.}\BibitemShut {Stop}%
\bibitem [{Note4()}]{Note4}%
  \BibitemOpen
  \bibinfo {note} {The subindex $(1)$ refers to first descendant.}\BibitemShut
  {Stop}%
\bibitem [{\citenamefont {Kitaev}(2001)}]{Kitaev01}%
  \BibitemOpen
  \bibfield  {author} {\bibinfo {author} {\bibfnamefont {A.~Y.}\ \bibnamefont
  {Kitaev}},\ }\href {http://stacks.iop.org/1063-7869/44/i=10S/a=S29}
  {\bibfield  {journal} {\bibinfo  {journal} {Phys. Usp.}\ }\textbf {\bibinfo
  {volume} {44}},\ \bibinfo {pages} {131} (\bibinfo {year} {2001})}\BibitemShut
  {NoStop}%
\bibitem [{\citenamefont {Qi}\ \emph {et~al.}(2009)\citenamefont {Qi},
  \citenamefont {Hughes}, \citenamefont {Raghu},\ and\ \citenamefont
  {Zhang}}]{Qi09}%
  \BibitemOpen
  \bibfield  {author} {\bibinfo {author} {\bibfnamefont {X.-L.}\ \bibnamefont
  {Qi}}, \bibinfo {author} {\bibfnamefont {T.~L.}\ \bibnamefont {Hughes}},
  \bibinfo {author} {\bibfnamefont {S.}~\bibnamefont {Raghu}}, \ and\ \bibinfo
  {author} {\bibfnamefont {S.-C.}\ \bibnamefont {Zhang}},\ }\href {\doibase
  10.1103/PhysRevLett.102.187001} {\bibfield  {journal} {\bibinfo  {journal}
  {Phys. Rev. Lett.}\ }\textbf {\bibinfo {volume} {102}},\ \bibinfo {pages}
  {187001} (\bibinfo {year} {2009})}\BibitemShut {NoStop}%
\bibitem [{\citenamefont {Wang}\ \emph {et~al.}(2014)\citenamefont {Wang},
  \citenamefont {Xu},\ and\ \citenamefont {Zhang}}]{Wang14}%
  \BibitemOpen
  \bibfield  {author} {\bibinfo {author} {\bibfnamefont {J.}~\bibnamefont
  {Wang}}, \bibinfo {author} {\bibfnamefont {Y.}~\bibnamefont {Xu}}, \ and\
  \bibinfo {author} {\bibfnamefont {S.-C.}\ \bibnamefont {Zhang}},\ }\href
  {\doibase 10.1103/PhysRevB.90.054503} {\bibfield  {journal} {\bibinfo
  {journal} {Phys. Rev. B}\ }\textbf {\bibinfo {volume} {90}},\ \bibinfo
  {pages} {054503} (\bibinfo {year} {2014})}\BibitemShut {NoStop}%
\bibitem [{\citenamefont {Fu}\ \emph {et~al.}(2007)\citenamefont {Fu},
  \citenamefont {Kane},\ and\ \citenamefont {Mele}}]{Fu07_2}%
  \BibitemOpen
  \bibfield  {author} {\bibinfo {author} {\bibfnamefont {L.}~\bibnamefont
  {Fu}}, \bibinfo {author} {\bibfnamefont {C.~L.}\ \bibnamefont {Kane}}, \ and\
  \bibinfo {author} {\bibfnamefont {E.~J.}\ \bibnamefont {Mele}},\ }\href
  {\doibase 10.1103/PhysRevLett.98.106803} {\bibfield  {journal} {\bibinfo
  {journal} {Phys. Rev. Lett.}\ }\textbf {\bibinfo {volume} {98}},\ \bibinfo
  {pages} {106803} (\bibinfo {year} {2007})}\BibitemShut {NoStop}%
\bibitem [{\citenamefont {Zhang}\ \emph {et~al.}(2009)\citenamefont {Zhang},
  \citenamefont {Liu}, \citenamefont {Qi}, \citenamefont {Dai}, \citenamefont
  {Fang},\ and\ \citenamefont {Zhang}}]{Zhang09}%
  \BibitemOpen
  \bibfield  {author} {\bibinfo {author} {\bibfnamefont {H.}~\bibnamefont
  {Zhang}}, \bibinfo {author} {\bibfnamefont {C.-X.}\ \bibnamefont {Liu}},
  \bibinfo {author} {\bibfnamefont {X.-L.}\ \bibnamefont {Qi}}, \bibinfo
  {author} {\bibfnamefont {X.}~\bibnamefont {Dai}}, \bibinfo {author}
  {\bibfnamefont {Z.}~\bibnamefont {Fang}}, \ and\ \bibinfo {author}
  {\bibfnamefont {S.-C.}\ \bibnamefont {Zhang}},\ }\href {\doibase
  10.1038/nphys1270} {\bibfield  {journal} {\bibinfo  {journal} {Nat. Phys.}\
  }\textbf {\bibinfo {volume} {5}},\ \bibinfo {pages} {438} (\bibinfo {year}
  {2009})}\BibitemShut {NoStop}%
\bibitem [{\citenamefont {Liu}\ \emph {et~al.}(2010)\citenamefont {Liu},
  \citenamefont {Qi}, \citenamefont {Zhang}, \citenamefont {Dai}, \citenamefont
  {Fang},\ and\ \citenamefont {Zhang}}]{Liu10}%
  \BibitemOpen
  \bibfield  {author} {\bibinfo {author} {\bibfnamefont {C.-X.}\ \bibnamefont
  {Liu}}, \bibinfo {author} {\bibfnamefont {X.-L.}\ \bibnamefont {Qi}},
  \bibinfo {author} {\bibfnamefont {H.}~\bibnamefont {Zhang}}, \bibinfo
  {author} {\bibfnamefont {X.}~\bibnamefont {Dai}}, \bibinfo {author}
  {\bibfnamefont {Z.}~\bibnamefont {Fang}}, \ and\ \bibinfo {author}
  {\bibfnamefont {S.-C.}\ \bibnamefont {Zhang}},\ }\href {\doibase
  10.1103/PhysRevB.82.045122} {\bibfield  {journal} {\bibinfo  {journal} {Phys.
  Rev. B}\ }\textbf {\bibinfo {volume} {82}},\ \bibinfo {pages} {045122}
  (\bibinfo {year} {2010})}\BibitemShut {NoStop}%
\bibitem [{\citenamefont {Kane}\ and\ \citenamefont
  {Mele}(2005{\natexlab{a}})}]{Kane05}%
  \BibitemOpen
  \bibfield  {author} {\bibinfo {author} {\bibfnamefont {C.~L.}\ \bibnamefont
  {Kane}}\ and\ \bibinfo {author} {\bibfnamefont {E.~J.}\ \bibnamefont
  {Mele}},\ }\href {\doibase 10.1103/PhysRevLett.95.146802} {\bibfield
  {journal} {\bibinfo  {journal} {Phys. Rev. Lett.}\ }\textbf {\bibinfo
  {volume} {95}},\ \bibinfo {pages} {146802} (\bibinfo {year}
  {2005}{\natexlab{a}})}\BibitemShut {NoStop}%
\bibitem [{\citenamefont {Kane}\ and\ \citenamefont
  {Mele}(2005{\natexlab{b}})}]{Kane05_2}%
  \BibitemOpen
  \bibfield  {author} {\bibinfo {author} {\bibfnamefont {C.~L.}\ \bibnamefont
  {Kane}}\ and\ \bibinfo {author} {\bibfnamefont {E.~J.}\ \bibnamefont
  {Mele}},\ }\href {\doibase 10.1103/PhysRevLett.95.226801} {\bibfield
  {journal} {\bibinfo  {journal} {Phys. Rev. Lett.}\ }\textbf {\bibinfo
  {volume} {95}},\ \bibinfo {pages} {226801} (\bibinfo {year}
  {2005}{\natexlab{b}})}\BibitemShut {NoStop}%
\bibitem [{\citenamefont {Bernevig}\ \emph {et~al.}(2006)\citenamefont
  {Bernevig}, \citenamefont {Hughes},\ and\ \citenamefont
  {Zhang}}]{Bernevig06}%
  \BibitemOpen
  \bibfield  {author} {\bibinfo {author} {\bibfnamefont {B.~A.}\ \bibnamefont
  {Bernevig}}, \bibinfo {author} {\bibfnamefont {T.~L.}\ \bibnamefont
  {Hughes}}, \ and\ \bibinfo {author} {\bibfnamefont {S.-C.}\ \bibnamefont
  {Zhang}},\ }\href {\doibase 10.1126/science.1133734} {\bibfield  {journal}
  {\bibinfo  {journal} {Science}\ }\textbf {\bibinfo {volume} {314}},\ \bibinfo
  {pages} {1757} (\bibinfo {year} {2006})}\BibitemShut {NoStop}%
\bibitem [{\citenamefont {Bernevig}\ and\ \citenamefont
  {Zhang}(2006)}]{Bernevig06_2}%
  \BibitemOpen
  \bibfield  {author} {\bibinfo {author} {\bibfnamefont {B.~A.}\ \bibnamefont
  {Bernevig}}\ and\ \bibinfo {author} {\bibfnamefont {S.-C.}\ \bibnamefont
  {Zhang}},\ }\href {\doibase 10.1103/PhysRevLett.96.106802} {\bibfield
  {journal} {\bibinfo  {journal} {Phys. Rev. Lett.}\ }\textbf {\bibinfo
  {volume} {96}},\ \bibinfo {pages} {106802} (\bibinfo {year}
  {2006})}\BibitemShut {NoStop}%
\bibitem [{\citenamefont {Fu}\ and\ \citenamefont {Kane}(2007)}]{Fu07}%
  \BibitemOpen
  \bibfield  {author} {\bibinfo {author} {\bibfnamefont {L.}~\bibnamefont
  {Fu}}\ and\ \bibinfo {author} {\bibfnamefont {C.~L.}\ \bibnamefont {Kane}},\
  }\href {\doibase 10.1103/PhysRevB.76.045302} {\bibfield  {journal} {\bibinfo
  {journal} {Phys. Rev. B}\ }\textbf {\bibinfo {volume} {76}},\ \bibinfo
  {pages} {045302} (\bibinfo {year} {2007})}\BibitemShut {NoStop}%
\bibitem [{\citenamefont {Moore}\ and\ \citenamefont
  {Balents}(2007)}]{Moore07}%
  \BibitemOpen
  \bibfield  {author} {\bibinfo {author} {\bibfnamefont {J.~E.}\ \bibnamefont
  {Moore}}\ and\ \bibinfo {author} {\bibfnamefont {L.}~\bibnamefont
  {Balents}},\ }\href {\doibase 10.1103/PhysRevB.75.121306} {\bibfield
  {journal} {\bibinfo  {journal} {Phys. Rev. B}\ }\textbf {\bibinfo {volume}
  {75}},\ \bibinfo {pages} {121306} (\bibinfo {year} {2007})}\BibitemShut
  {NoStop}%
\bibitem [{\citenamefont {Wang}\ \emph {et~al.}(2010)\citenamefont {Wang},
  \citenamefont {Qi},\ and\ \citenamefont {Zhang}}]{Wang_2010}%
  \BibitemOpen
  \bibfield  {author} {\bibinfo {author} {\bibfnamefont {Z.}~\bibnamefont
  {Wang}}, \bibinfo {author} {\bibfnamefont {X.-L.}\ \bibnamefont {Qi}}, \ and\
  \bibinfo {author} {\bibfnamefont {S.-C.}\ \bibnamefont {Zhang}},\ }\href
  {\doibase 10.1088/1367-2630/12/6/065007} {\bibfield  {journal} {\bibinfo
  {journal} {New Journal of Physics}\ }\textbf {\bibinfo {volume} {12}},\
  \bibinfo {pages} {065007} (\bibinfo {year} {2010})}\BibitemShut {NoStop}%
\bibitem [{\citenamefont {Yakovenko}(1990)}]{Yakovenko90}%
  \BibitemOpen
  \bibfield  {author} {\bibinfo {author} {\bibfnamefont {V.~M.}\ \bibnamefont
  {Yakovenko}},\ }\href {\doibase 10.1103/PhysRevLett.65.251} {\bibfield
  {journal} {\bibinfo  {journal} {Phys. Rev. Lett.}\ }\textbf {\bibinfo
  {volume} {65}},\ \bibinfo {pages} {251} (\bibinfo {year} {1990})}\BibitemShut
  {NoStop}%
\bibitem [{\citenamefont {Hlousek}\ \emph {et~al.}(1990)\citenamefont
  {Hlousek}, \citenamefont {S\'en\'echal},\ and\ \citenamefont
  {Tye}}]{Hlousek90}%
  \BibitemOpen
  \bibfield  {author} {\bibinfo {author} {\bibfnamefont {Z.}~\bibnamefont
  {Hlousek}}, \bibinfo {author} {\bibfnamefont {D.}~\bibnamefont
  {S\'en\'echal}}, \ and\ \bibinfo {author} {\bibfnamefont {S.~H.~H.}\
  \bibnamefont {Tye}},\ }\href {\doibase 10.1103/PhysRevD.41.3773} {\bibfield
  {journal} {\bibinfo  {journal} {Phys. Rev. D}\ }\textbf {\bibinfo {volume}
  {41}},\ \bibinfo {pages} {3773} (\bibinfo {year} {1990})}\BibitemShut
  {NoStop}%
\bibitem [{\citenamefont {Gurarie}(2011)}]{Gurarie11}%
  \BibitemOpen
  \bibfield  {author} {\bibinfo {author} {\bibfnamefont {V.}~\bibnamefont
  {Gurarie}},\ }\href {\doibase 10.1103/PhysRevB.83.085426} {\bibfield
  {journal} {\bibinfo  {journal} {Phys. Rev. B}\ }\textbf {\bibinfo {volume}
  {83}},\ \bibinfo {pages} {085426} (\bibinfo {year} {2011})}\BibitemShut
  {NoStop}%
\bibitem [{\citenamefont {Santos}\ \emph {et~al.}(2011)\citenamefont {Santos},
  \citenamefont {Nishida}, \citenamefont {Chamon},\ and\ \citenamefont
  {Mudry}}]{Santos11}%
  \BibitemOpen
  \bibfield  {author} {\bibinfo {author} {\bibfnamefont {L.}~\bibnamefont
  {Santos}}, \bibinfo {author} {\bibfnamefont {Y.}~\bibnamefont {Nishida}},
  \bibinfo {author} {\bibfnamefont {C.}~\bibnamefont {Chamon}}, \ and\ \bibinfo
  {author} {\bibfnamefont {C.}~\bibnamefont {Mudry}},\ }\href {\doibase
  10.1103/PhysRevB.83.104522} {\bibfield  {journal} {\bibinfo  {journal} {Phys.
  Rev. B}\ }\textbf {\bibinfo {volume} {83}},\ \bibinfo {pages} {104522}
  (\bibinfo {year} {2011})}\BibitemShut {NoStop}%
\bibitem [{\citenamefont {Chen}(2018)}]{Chen18}%
  \BibitemOpen
  \bibfield  {author} {\bibinfo {author} {\bibfnamefont {W.}~\bibnamefont
  {Chen}},\ }\href {\doibase 10.1103/PhysRevB.97.115130} {\bibfield  {journal}
  {\bibinfo  {journal} {Phys. Rev. B}\ }\textbf {\bibinfo {volume} {97}},\
  \bibinfo {pages} {115130} (\bibinfo {year} {2018})}\BibitemShut {NoStop}%
\end{thebibliography}%

\end{document}